\documentclass[%
 reprint,
 amsmath,amssymb,
 aps,
]{revtex4-2}
\usepackage{graphicx}
\usepackage{dcolumn}
\usepackage{color}
\usepackage{bm}
\usepackage{bbm}
\usepackage{blindtext}
\usepackage{float}
\usepackage{hyperref}
\usepackage{amsmath}
\usepackage{tikz}
\usepackage{aas_macros}
\usepackage[normalem]{ulem}
\hypersetup{
    colorlinks=true,
    linkcolor=blue,
    filecolor=magenta,      
    urlcolor=cyan,
    pdftitle={Overleaf Example},
    pdfpagemode=FullScreen,
    }

\begin{document}

\preprint{APS/123-QED}
\title{Mitigating the effects of instrumental artifacts on source localizations}
\author{Maggie C. Huber$^{1,2}$ and
        Derek Davis$^3$
        }
\affiliation{$^1$University of Michigan, Ann Arbor, MI 48104, USA\\
$^2$University of Colorado Boulder, Boulder, CO 80309, USA\\
$^3$LIGO, California Institute of Technology, Pasadena, CA 91125, USA}
\date{\today}

\begin{abstract}

Instrumental artifacts in gravitational-wave strain data can overlap with gravitational-wave detections and significantly impair the accuracy of the measured source localizations. These biases can prevent the detection of any electromagnetic counterparts to the detected gravitational wave. We present a method to mitigate the effect of instrumental artifacts on the measured source localization. This method uses inpainting techniques to remove data containing the instrumental artifact and then correcting for the data removal in the subsequent analysis of the data.  
We present a series of simulations using this method using a variety of signal classes and inpainting parameters which test the effectiveness of this method and identify potential limitations. 
We show that in the vast majority of scenarios, this method can robustly localize gravitational-wave signals even after removing portions of the data. 
We also demonstrate how an instrumental artifact can bias the measured source location and how this method can be used to mitigate this bias.
\end{abstract}

\maketitle

\section{Introduction}
Modern gravitational-wave (GW) detectors including Advanced LIGO~\cite{LIGOScientific:2014pky}, Advanced Virgo~\cite{TheVirgo:2014hva}, and KAGRA~\cite{KAGRA:2018plz} are increasingly sensitive to compact-binary coalescences (CBCs)~\cite{GWTC-1,GWTC-2,GWTC-3,GWTC-2.1}. Binary neutron star (BNS) and neutron star-black hole (NSBH) mergers are of particular interest~\cite{PhysRevLett.119.161101,2020ApJ...892L...3A,2021ApJ...915L...5A} as they are known to produce a variety of transient electromagnetic (EM) signals across multiple wavebands. Relativistic jets from BNS mergers can be progenitors to short gamma-ray bursts (GRBs)~\cite{2017ApJ...848L..13A,2017ApJ...848L..14G} which in turn cause afterglow radiation in X-ray, optical, and radio bands lasting from hours to years~\cite{doi:10.1126/science.aap9855,2017ApJ...848L..21A,2017Natur.551...71T,2018NatAs...2..751L, Balasubramanian:2021kny}. Ejected material from BNS mergers also produces kilonovae observable in the optical and near-infrared band for days to weeks~\cite{Metzger:2019zeh,2017ApJ...848L..19C,2017Sci...358.1559K}. EM followup campaigns are crucial for advancing the field of multi-messenger astrophysics~\cite{LIGOScientific:2017ync}. These events have allowed us to test general relativity~\cite{PhysRevD.103.122002}, characterize the nuclear equation of state~\cite{LIGOScientific:2018cki} and the black hole population~\cite{2021arXiv211103634T}, and discover objects that challenge existing models for stellar evolution~\cite{abbott2020gw190521,2020ApJ...896L..44A}. 

In order for telescopes to observe EM radiation from BNS mergers, the GW detector network must triangulate and localize the GW source to produce a  skymap~\cite{2011CQGra..28j5021F,2016ApJ...829L..15S}. These skymaps tend to cover a large portion of the sky compared to most telescopes' field of view~\cite{2018LRR....21....3A}. This can make sources difficult to rapidly localize and make EM followup challenging to optimize\cite{Coughlin:2018lta,Coughlin:2019qkn,PhysRevD.81.082001,2013ApJ...767..124N,PhysRevD.89.084060}. Generating rapid and accurate skymaps is important for the success of EM followup campaigns and helps reveal useful astrophysical information about a GW event.

Skymap accuracy and other estimates of GW source parameters can be impaired by the presence of noise from instrumental artifacts in GW detectors~\cite{Macas:2022afm,powell2018parameter,Cabero:2020eik,Vitale:2016jlv}. One main class of instrumental artifact is ``glitches,'' which manifest as bursts of excess power~\cite{LIGO:2021ppb,aasi2012characterization,KAGRA:2020agh}. Often, what causes these glitches is difficult to determine. They can be 
the result of either external environmental or internal instrumental interactions that alter the actual GW strain~\cite{nguyen2021environmental,2018RSPTA.37670286N,davis2021ligo,2018JPhCS.957a2004B}. Glitches are particularly troublesome for multi-messenger observations, as the are more likely to overlap with GW events with a longer duration, such as BNS and NSBH mergers. This has already occurred seen in all previously-detected confident BNS and NSBH mergers~\cite{PhysRevLett.119.161101,2020ApJ...892L...3A,2021ApJ...915L...5A}. As detection of GW events from CBC mergers becomes increasingly frequent, 
we expect to see more instances of glitches overlapping with GW signals. 

There are multiple approaches one can use to address a glitch which overlaps a GW signal. In the case of GW170817, the effects of a glitch overlapping the signal were mitigated by applying a window function to remove the data containing the glitch~\cite{Pankow:2018qpo,harris1978use}. The glitch waveform was then later reconstructed using wavelets
\cite{Cornish:2020dwh,cornish2015bayeswave,cornish2021bayeswave} that could be subtracted from the data 
\cite{PhysRevLett.119.161101}. Although modelling glitches has been successfully used in previous GW analyses~\cite{PhysRevLett.119.161101,GWTC-2, GWTC-3, GWTC-2.1}, this method can be slow and ad hoc in nature, as we cannot always get an accurate glitch model. Different approaches are necessary to find a generalized solution that works rapidly for various types of glitches. There have already been efforts to address this issue using machine learning techniques~\cite{Mogushi:2021cpw,cuoco2020enhancing}. However, these algorithms must be tuned to specific waveforms and have only been tested for very short glitch durations.

\begin{figure}
    \centering
    \vspace{0.5cm}
    \includegraphics[width=\columnwidth]{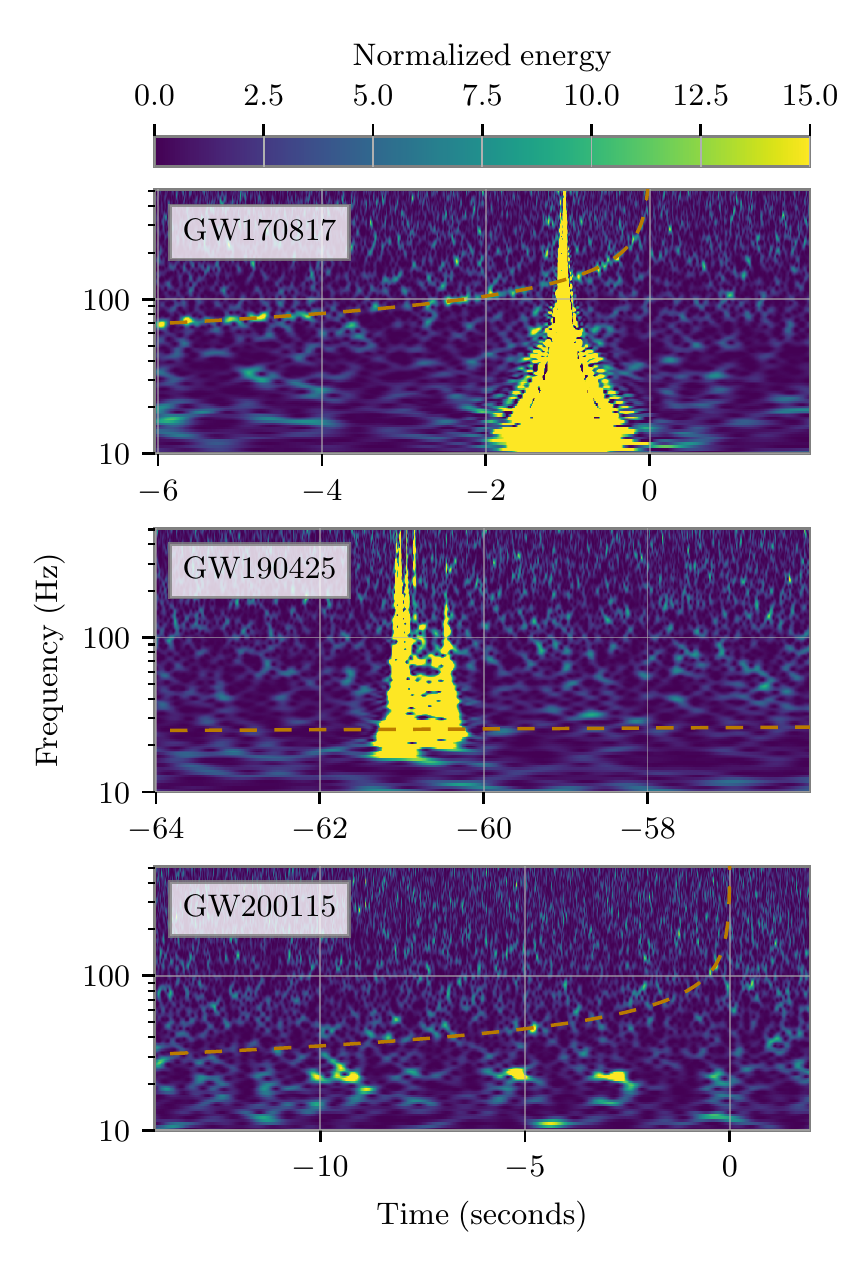}
    \caption{Time-frequency plot of LIGO Livingston data produced with with Q-transform~\cite{Chatterji:2004qg} showing glitches overlapping with GW170817~\cite{TheLIGOScientific:2017qsa}, GW190425~\cite{2020ApJ...892L...3A}, and GW200115\_042309 (GW200115)~\cite{LIGOScientific:2021qlt}. The time-frequency tracks of the gravitational-wave signals are shown as orange dotted lines. This figure illustrates how bursts of excess power interfere with time-frequency data and obscure the merger.} 
    \label{fig1}
    
\end{figure}

When editing the recorded strain data to mitigate a glitch, care must be taken to minimize the potential biases the mitigation procedure introduces. Window functions such as the one used for GW170817 add a taper to the edges of the removed data to avoid introducing discontinuities. However, they can introduce excess power leakage from the spectral lines in the power spectral density (PSD) of the detector. An alternative to window functions is inpainting 
\cite{2019arXiv190805644Z}, where the effects of discontinuities are calculated and subtracted. The end result is a gate that only masks bad seconds of data and should have no effect on the data surrounding the inpainted hole.

Inpainting, however, is not without its limitations. 
When we inpaint a hole in GW data we lose information about the amplitude and phase of a signal. This in turn biases the sky localization, although it is less noticeable when the inpainted duration is significantly less than the total signal. For larger inpainting widths, this can add a significant bias and make skymaps less useful for localizing sources. To ensure a skymap that is accurate and precise, we must correct for the effect of gating a portion of the signal.

To effectively mitigate glitches and reduce bias in our skymaps, we developed an algorithm to inpaint, reweight, and normalize the signal-to-noise ratio (SNR) time series of the signal. In this paper we explain the functionality of this new glitch mitigation method, the setup of our Python package, Python-based  Skymap  Localization with Inpainted Data Editor (pySLIDE), and the metrics we used to show how well it performs. Our work combines techniques to manage glitches into a simple and computationally efficient package, and rigorously tests how our methods perform. The goal is to recover the correct error and accuracy so that the skymap includes the signal and guides EM telescopes in the right direction.  We determine that our method was able to recover a more accurate skymaps both in the case of simulated signals with and without a glitch for a variety of GW source parameters. Additionally, the package is computationally efficient and has a deterministic underlying formula with flexible input parameters.

In Section~\ref{sec:BAYESTAR}, we explain the relevant components of BAYESTAR sky localization. In~\ref{sec:Signal}, we go into the processes for calculating the SNR time series and obtaining results. In~\ref{sec:Reweighting}, we discuss the  reweighting and normalizing technique we developed for this method. In~\ref{sec:PySLIDE Workflow}, we touch on the setup of pySLIDE. In~\ref{sec:Testing simulated signals}, we describe our tests of the method. In Sections~\ref{sec:validation} and~\ref{sec:glitch}, we present our results and in Section~\ref{sec:Conclusions} we discuss important takeaways from this research.

\section{Skymap computation}
\subsection{BAYESTAR}
\label{sec:BAYESTAR}

We used the BAYESTAR~\cite{Singer_2016} algorithm and the PyCBC search pipeline~\cite{Usman:2015kfa,nitz2018rapid} to create our sky localizations. BAYESTAR localizes GW sources using Bayesian inference instead of Markov chain Monte Carlo (MCMC) methods. It takes a likelihood function and a well-defined parameter space to rapidly infer the location of GW signals.
Specifically, BAYESTAR requires information about the amplitude, the phase, and the relative time of arrival of the signal in all detectors. 

The BAYESTAR algorithm is used to rapidly localize gravitational-wave signals with the information that is available in low latency. 
To calculate the sky localization posterior, BAYESTAR uses the maximum likelihood values of the intrinsic source parameters, masses and spins, from a low latency matched filter search pipeline.
The BAYESTAR likelihood for each point in the sky is marginalized over additional extrinsic parameters; coalescence phase, absolute time of arrival, distance, inclination angle, and polarization~\cite{Singer_2016}:
\begin{equation}
f(\text{RA}, \text{Dec}) \propto
    \idotsint \mathcal{L}_{\text{BS}} \,r^3 \,d\phi_c\,dr\,dt\,d\cos{i}\,d\psi
\end{equation}
The BAYESTAR likelihood can be further segmented into a term based only on the SNR of the signal in each detector and a term based on the relative time of arrival (TOA) in each detector~\cite{Singer_2016}, 
\begin{equation}
\begin{split}
    \mathcal{L}_{\text{BS}} & \propto \mathcal{L}_{\text{SNR}} \times \mathcal{L}_{\text{TOA}} \\
    & \propto \exp{\left[-\frac{1}{2}\sum_i \rho_i^2 + 
        \sum_i \rho_i \mathbb{R}\{e^{-i\theta_i} z_i^*(\tau_i)\}\right]}
\end{split}
\end{equation}
where $\rho_i$ is the SNR, $\theta_i$ is the phase, and $z_i(\tau_i)$ is the SNR time series of the maximum likelihood template for a given detector, $i$.
The values for $\rho_i$ and $\theta_i$ are extracted from the SNR time series at the time of maximum SNR in each detector, $\tau_{i,max}$, by
\begin{equation}
\begin{split}
    \rho_i & = \left| z_i(\tau_{i,max}) \right| \\
    \theta_i & = \arg \left\{ z_i(\tau_{i,max}) \right\}
\end{split}
\end{equation}
An additional normalization term is also included the likelihood, which is dependent on the PSD of each detector. 
The BAYESTAR likelihood, and hence the sky localization posterior, is entirely dependent on the provided SNR time series and the PSD.
It is therefore these two products that we wish to modify as part of our glitch mitigation method. 

\subsection{Signal parameters}
\label{sec:Signal}

To evaluate these parameters for a gravitational-wave signal, we first use GWpy~\cite{gwpy} to generate a PSD and get a matched filter from PyCBC~\cite{Usman:2015kfa} to run the waveform template through the noisy data and calculate the SNR. 
The matched filter function for SNR time series $z(t)$ is given by a weighted inner product of the detector data $s$ and template $h(t)$. 
The SNR time series, $z(t)$, is written as~\cite{Allen:2005fk}
\begin{equation}
z(t) = \frac{(s|h)}{(h|h)^{1/2}}
    \text{ .}
\end{equation}
The inner product $(s|h)$ is given by 
\begin{equation}
    (s|h)(t) = 4\int_{f_{low}}^{f_{high}}\frac{\tilde{s}(f)\tilde{h^*}(f)}{S_{n}(f)}\textup{e}^{2\pi ift}\textup{d}f
\end{equation}
with $S_{n}(f)$ as the PSD. This is similar to convolving the template against the overwhitened data. The real part of this SNR time series corresponds to a template that is lined up along the data and the imaginary part corresponding to a template that is 90 degrees out of phase. The amplitude of the phase of the SNR time series can be then extracted from he amplitude and phase of the corresponding complex SNR. 

As described int he previous section, the amplitude of the SNR time series given by the matched filtering process is a key input into a sky localization for a GW signal. The other two parameters needed for a skymap are the time delay between each detector and the phase of the signal. We do not expect the presence of a glitch to bias the time delay measurements or the phase, so PySLIDE focuses on correcting the amplitude of the SNR time series. 

Once the localization parameters are input into BAYESTAR, we can extract important information
from the skymap it returns. 
Here we measure the credible regions and searched area to evaluate the performance of BAYESTAR. The credible region represents the cumulative sum of pixels in a given region. 
We are particularly interested in the 90 percent credible region, which contains 90 percent of pixels in the skymap. The smaller the area of this region, the more precise our skymap is. We also measure the total searched area, which is the smallest credible region containing the location of the source. A smaller total searched area corresponds to better accuracy.

\subsection{Inpainting and reweighting}
\label{sec:Reweighting}

\begin{figure*}[ht]
\centering
\includegraphics[width=\textwidth]{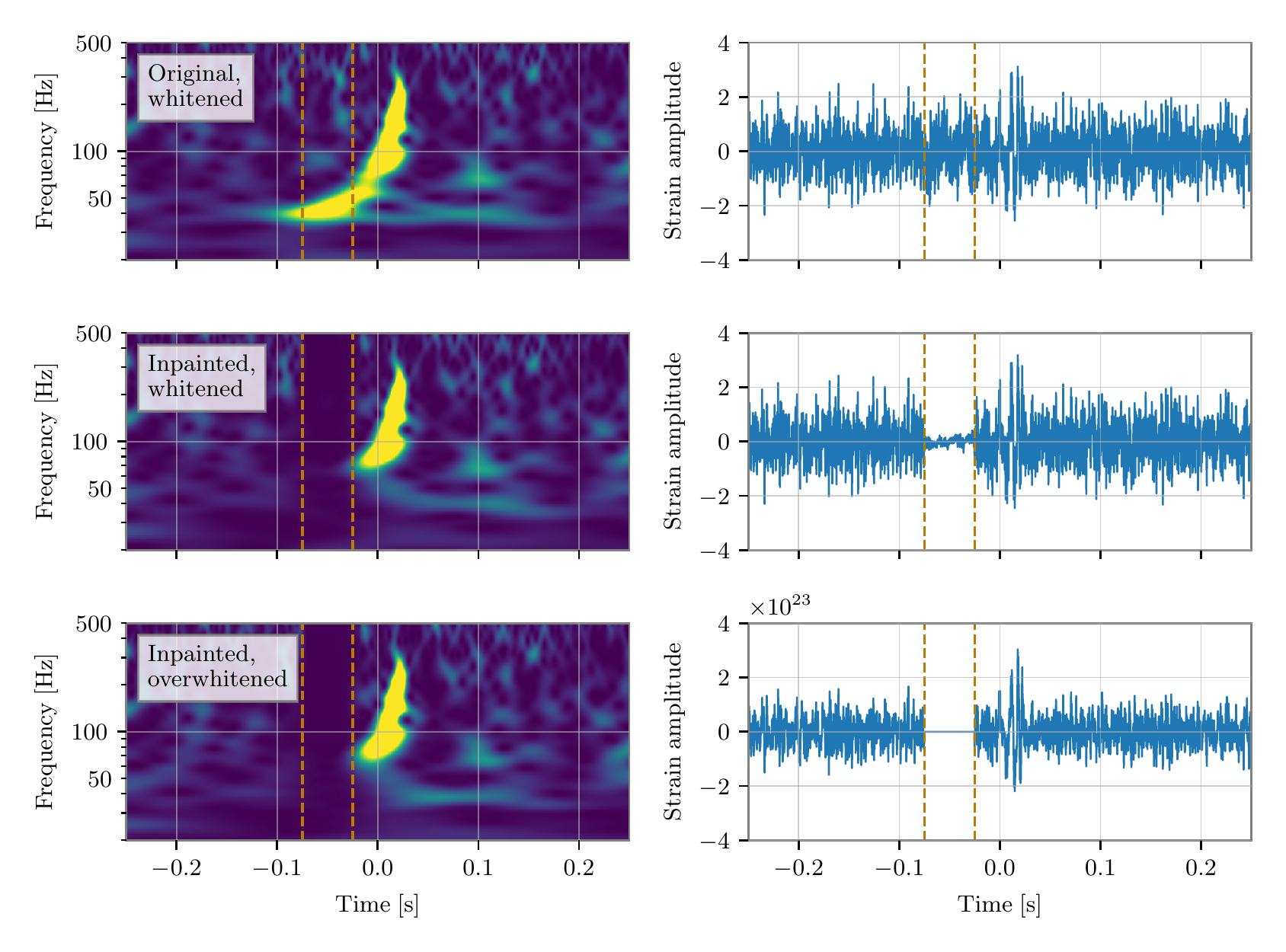}
\caption{A comparison of data surrounding GW150914~\cite{LIGOScientific:2016aoc} before and after inpainting. The inapainted data is shown both whitened and overwhitened. The left panels show spectrograms of the data while the right panels show the same data as a time series. The start and end times of the window that is inpainted is shown in each panel with dotted orange lines. Per the definition of the inpainting filter~\cite{2019arXiv190805644Z}, the data inside the inpainted region is only zeroed in the overwhitened data. }
\label{fig:inpaint}
\end{figure*}

If a glitch is present in the data, the SNR time series described in the previous section may deviate from the expected in Gaussian noise. 
In order to correct for this potential bias, we remove the data containing a glitch using a technique called inpainting~\cite{2019arXiv190805644Z} and recalculate the SNR time series.
The inpainting filter is designed such that the overwhitened, inpainted data is zeroed inside a time window of interest.
An example of how inpainting affects both whitened and overwhitened data is shown in Figure~\ref{fig:inpaint}.
The removal of data ensures that any data quality issues present in the data do not bias the SNR time series.
However, this data removal introduces it's own bias.
The benefit of inpainting lies in the fact that this bias is well known and can be corrected for in our analysis. 

To correct for the bias of inpainting, we first calculate the expected loss in SNR from the inpainting procedure. The 
fraction of the SNR remaining after inpainting is given by~\cite{2019arXiv190805644Z}

\begin{equation} \label{eq:1}
    \lambda_{hole}(t_{0},h)\approx \frac{(\left |h_{w}\right |^{2}\circledast\mathbbm{1}_{valid})(t_0)} {\sum_{t}\left |h_{w}(t)\right |^{2}} 
    \text{ ,}
\end{equation}
where $t_{0}$ is the merger time, $h_{w}$ is the whitened waveform, and $\mathbbm{1}_{valid}$ returns zero for a data point in the inpainted hole and one otherwise. The equation convolves $h_{w}$ with $\mathbbm{1}_{valid}$, which we compute with a fast Fourier transform (FFT) as allowed by the convolution theorem. When we use Eq. \eqref{eq:1} to calculate the normalization at each time, we can find the peak in the renormalized SNR time series $t_0$ by calculating the time of the maximum renormalized SNR.

After we inpaint and apply Equation~\ref{eq:1}, we multiply a normalization factor to the PSD and SNR time series using

\begin{equation} \label{eq:3}
    \tilde{z}(t) = z(t)\frac{\lambda(\tau_{i,max})}{\lambda(t)} \\
\end{equation}    \
\begin{equation}
    \tilde{S_n}(f) = \frac{S_n(f)}{\lambda(\tau_{i,max})} \\
    \text{ .}
\end{equation}
These new values are then used in BAYESTAR to recompute the skymap and obtain the credible region and searched areas for the renormalized time series.

When a portion of a signal is inpainted, we effectively decrease the sensitivity of our measurement. Renormalizing the PSD corrects the error for our BAYESTAR localization, and as a result we can avoid overestimating how much we know about the signal. 
If we inpaint a hole, we are losing information and the error should increase to account for that if we want the true source location to be included in the credible region of the skymap. This is different than restoring the SNR to where it was before removing data, which would bias our results in the other direction.

There are multiple advantages of using the reweighting method to renormalize the PSD. The algorithm can be used regardless 
of how much time is removed with inpainting and which waveform template we use, and it allows the user to configure these settings as inputs. Reweighting is 
also deterministic - the calculation is the same for any variation of the input parameters. It is also efficient 
to compute, typically taking less than a second. These benefits render this method conducive to rapid and accurate sky localization of GW events in real time, even in the presence of glitches.
\subsection{PySLIDE Workflow}
\label{sec:PySLIDE Workflow}
PySLIDE can apply and test the reweighting method to any number of signal injections, though there are diminishing returns when you get to about 500 injection runs. It creates a PyCondor~\cite{pycondor} workflow  that separates each task into jobs, which are then collected into a ``Dag''~\cite{pycondor} object to submit to a computing cluster.

There are several components of the package that create results and plot them. The first script applies the reweighting method to the SNR time series using 
injected signal parameters and a PSD. It creates a time series with the original data, inpaints a hole for a specified segment, applies the reweighting formula to the SNR and PSD, and corrects for the remaining SNR . The script outputs the SNR time series as an XML file to be fed into the BAYESTAR algorithm. 

BAYESTAR returns a localization as a FITS file. Using this file, we run scripts to calculate the credible region of the true source location, the total searched area, the area of the 90 percent credible region, and the overlap of the reweighted and inpainted skymaps with the original skymap~\cite{skymap-overlap}. PySLIDE combines the results from all injection runs into one file and then that file is used for four different plotting scripts to visualize the final results. This process is computationally efficient and takes less time to run than similar methods. BAYESTAR localization contributed to the majority of the computation time. 

The python package we developed can also be used to evaluate sky localization of candidates that are identified in real-time processing of gravitational wave data.
Either automated tools or visual inspection of the data can quickly be used to identify any glitches that overlap identified signals. 
These glitches can then be mitigated with the inpainting methods discussed in this work and the skymap can be recomputed using the same waveform template that was used to identify the signal. 
If the new skymap is different from the skymap produced before any glitches were mitigated, this could indicate a bias in the original skymap.

\section{Tests with simulated signals}
\label{sec:Results}

\subsection{Description of tests}
\label{sec:Testing simulated signals}

To assess the performance of our reweighting algorithm, we simulated various CBC signals for testing. For the waveform template, we selected SEOBNRv4\_ROM~\cite{Bohe:2016gbl} and filtered our simulated parameter list to include what we would expect to detect by applying an SNR threshold of 8. We varied the duration and offset of the gate from the merger time. We also ran our tests over different component masses corresponding to BNS, NSBH, and binary black hole (BBH) mergers.
Table~\ref{tab:config} lists the combinations of mass, inpainting window, and offset we tested in this work. 
Throughout the rest of this work, we refer to a specific combination of inpainting window and offset as ``(inpainting window in seconds, offset in seconds).''

To run our tests, we inject the simulated signals into Gaussian noise colored to representative PSDs from the LIGO Hanford (H1) and Livingston (L1) detectors and get the original SNR time series. We then calculate the SNR time series when using the inpainting function 
alone, then when inpainting and reweighting. We create an XML file which is put into BAYESTAR to localize all three cases. To see how the method performs, we calculate the credible regions, searched area, the area of the 90 percent credible region, and the overlap for all injected signals.

\begin{table}[]
\begin{tabular}{llll}
\multicolumn{1}{r}{Masses $(M_\odot)$}       & \begin{tabular}[c]{@{}l@{}}Mass 1 = 1.4 \\ Mass 2 = 1.4\end{tabular} & \begin{tabular}[c]{@{}l@{}}Mass 1 = 30  \\ Mass 2 = 1.4\end{tabular} & \begin{tabular}[c]{@{}l@{}}Mass 1 = 30 \\ Mass 2 = 30\end{tabular} \\ \hline
\multicolumn{1}{l|}{Duration (s)} & 0.5, 1, 4                                                            & 0.125, 0.5, 1                                                        & 0.0625, 0.125, 0.5                                                 \\
\multicolumn{1}{l|}{Offset (s)}   & 0, -1, -32                                                           & 0, -1, -4                                                            & 0, -0.1, -0.25                                                    
\end{tabular}
\caption{Summary of the tests we ran for various GW signals. We tried combinations of three different durations and offsets specific to each set of component masses. This corresponds to 9 total combinations per class of signal.}
\label{tab:config}
\end{table}

\subsection{Validation test results}
\label{sec:validation}

\begin{figure*}[ht]
\centering
\includegraphics[width=\textwidth]{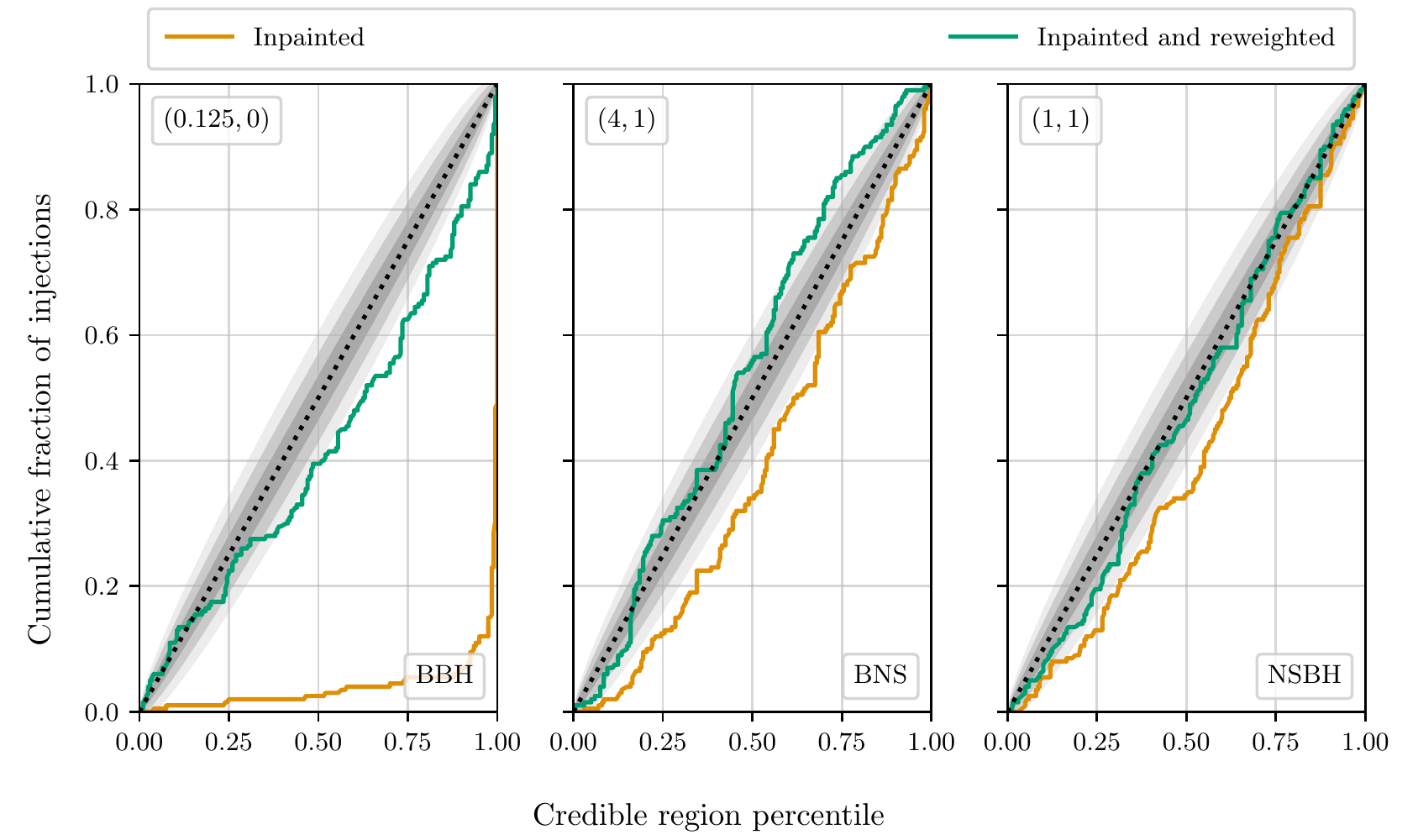}
\caption{P-P plot for three sets of simulations showing the distribution of the credible region returned by BAYESTAR of each injection's 
true source location. Shown is the distribution when the data is only inpainted (orange) and when the data is both inpainted and reweighted (green). The distributions are normalized by the measured distribution when no mitigation is used. The dotted line and the shaded regions show the expected distribution and the 1-, 2-, and 3-$\sigma$ uncertainties in the expected distribution. The (window size, offset from merger) for each test are shown in the upper left while the signal type is shown on the bottom right. For the leftmost panel with BBH distribution, we can see that inpainting significantly biases the P-P plot. After reweighting we see that the error improves. For the BNS distribution in the middle, inpainting similarly biases the results but it is not as large of an effect. The NSBH distribution on the right does not show as drastic of a change, but gets an accurate error recovery.}
\label{fig2}
\end{figure*}

To understand if the previously described method is accurately and precisely recovering the sky location of our injected signal, we focus on three metrics:
the fraction of signals recovered within each credible region, the searched area, and the 90 percent credible area. 
Descriptions and results for each metric are explained later in this section.
In this section, we focus on three example configurations of inpainting window, offset, and signal class. 
The complete results for all tested configuration combinations are presented in Appendix~\ref{app:all_results}.

The first metric we investigate is whether the expected fraction of injections are recovered within a specific credible region. This metric is often referred to as a ``probability-probability'' (P-P) plot, and shows the credible region of the true source location vs. the fraction of total simulated signals that fall there (Figure~\ref{fig2}). These should match up, i.e. 90 percent of the signals we create end up in the 90 percent credible region of the skymap. On a P-P plot, this distribution is linear with a slope of one. Due to some complications including an internal fudge factor in BAYESTAR, our original data without the glitch lies above the diagonal when we use PyCBC. Additionally, whether the P-P plot lies on the one-to-one line depends on the component masses of our signal. We are accounting for these complications by normalizing the P-P plots by the original data for the simulated signals without a glitch. We are only comparing our results to the original data and the ideal credible region will be a one-to-one plot. Credible regions that lie above the line are overestimating error in the skymap, and regions under the line are underestimating error. 

From Figure~\ref{fig2}, we see that inpainting a hole in the data will bias the skymap and report an incorrect error. When we reweight the SNR time series, correctly sized credible regions are recovered and the skymap is more likely to return a localization that contains the source. How well reweighting corrects inpainting bias depends on the duration, offset, and signal type. For the BBH signals, Figure~\ref{fig2} shows an extreme case where the inpainted hole was centered on the merger and was a large portion of the signal length. This indicates that when we lose almost all information about the signal, the error increases accordingly to reflect that there is limited information we can extract about the signal. For the BNS case, when the inpainted hole is offset from the merger reweighting is able to get the credible regions closer to the original data and overestimating the error. For the NSBH component masses, we are also able to recover the correct credible regions and slightly underestimating the error. 
Despite the range of masses and inpainting configurations, we are able to approximately recover the correctly sized credible regions in all cases presented in this section.

\begin{figure*}[ht]
\centering
\includegraphics[width=\textwidth]{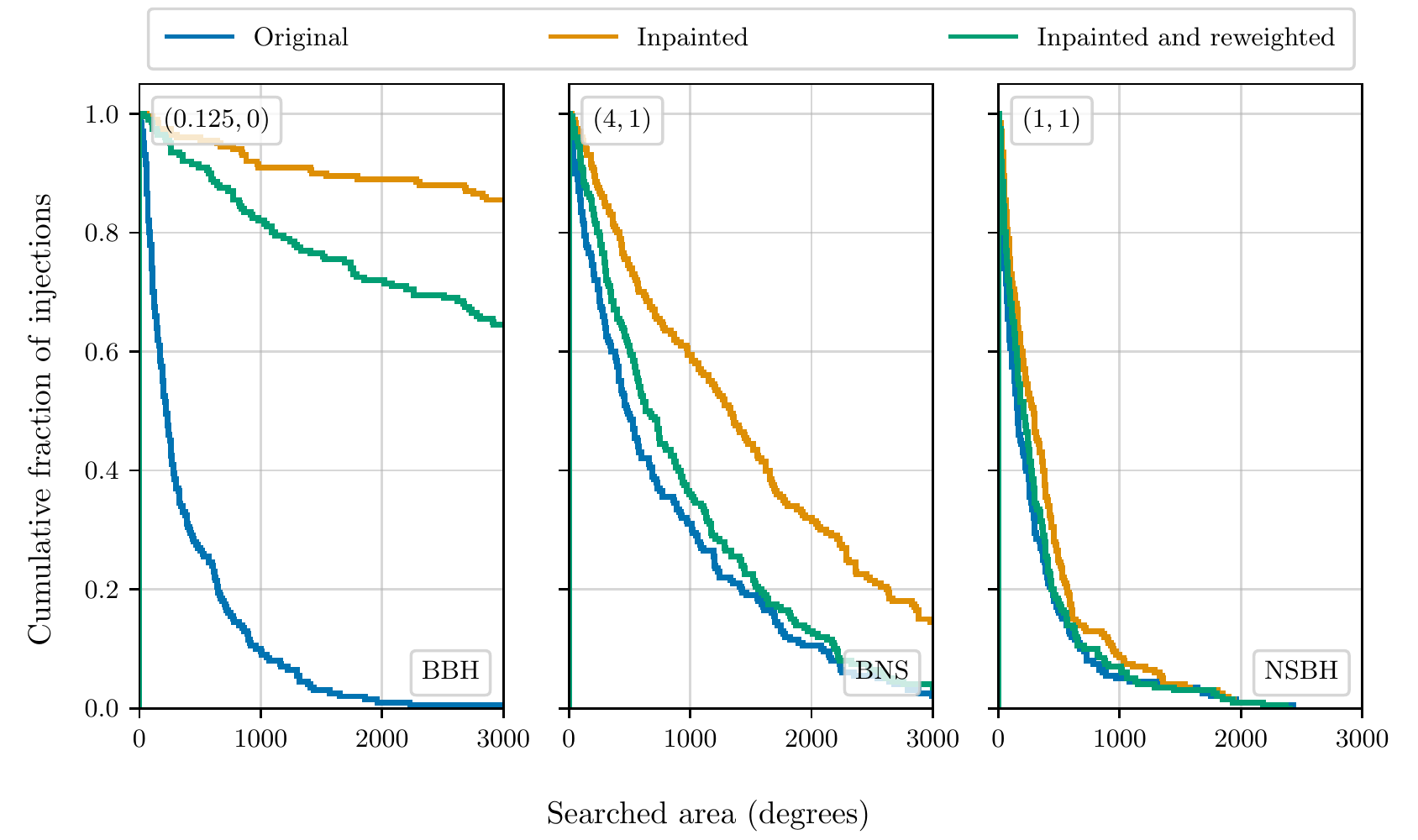}
\caption{Histogram showing the total searched area without a glitch by BAYESTAR in degrees vs. the cumulative sum of signals. The (window size, offset from merger) for each test are shown in the upper left while the signal type is shown on the bottom right. For all cases, the searched area improves after reweighting and recovers a searched area closer to the original before inpainting. For the BBH case, the searched area slightly improves, and we can see in the BNS and NSBH signals our accuracy lines up much more closely with the original.}
\label{fig3}
\end{figure*}

To check if the skymap shows an accurate credible region, we create a histogram of the total searched area, as shown in Figure~\ref{fig3}. The searched area is the area of the smallest credible region containing the true source location. Ideally, the cumulative fraction drops off faster as the searched area increases. This demonstrates that the resulting skymap predicts the source location to be in a lower credible region. For each signal and gate type, the amount of accuracy recovered after reweighting varies. Figure~\ref{fig3} shows that for these various signals and gates, the searched area is closer to the original and accuracy does improve after reweighting.
Although we would ideally recover the original searched area distribution after both inpainting and reweighting, it is expected that we would still have larger searched area due to the loss of information that inpainting introduces. 
This information loss naturally leads to less accurate sky maps. 
However, the inpainting and reweighted maps are more accurate than when only inpainting. 

\begin{figure*}[ht]
\centering
\includegraphics[width=\textwidth]{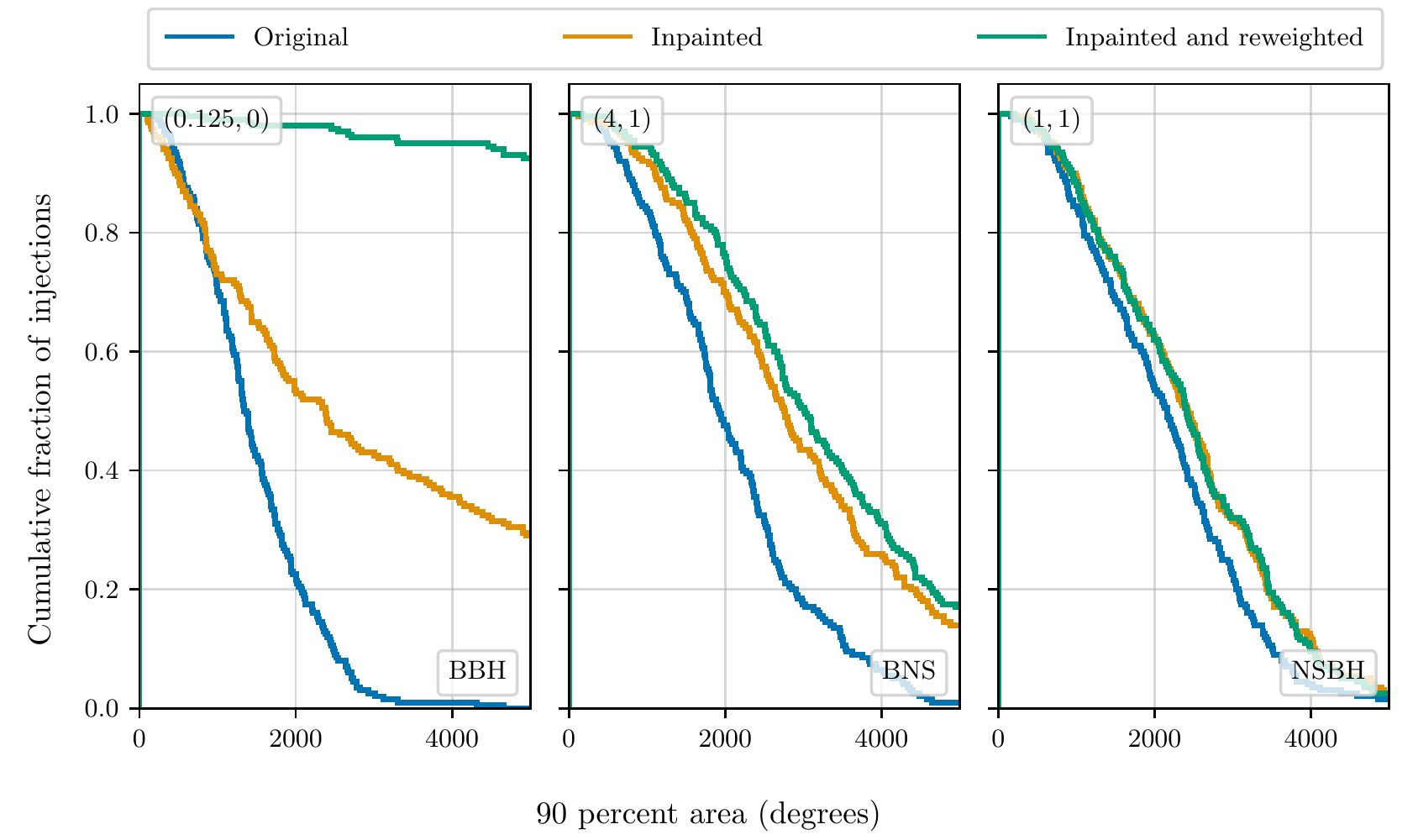}
\caption{Histogram showing the area of the 90 percent credible region on the skymap, which we use to roughly estimate the precision of our measurement. The (window size, offset from merger) for each test are shown in the upper left while the signal type is shown on the bottom right. We can see that the distribution of the 90 percent areas includes significantly larger values, particularly in the BBH signal where we have removed a large portion of the data and know less about our signal. For the BNS merger in the middle, the 90 percent credible region only increases slightly, and is roughly the same in the right panel with the NSBH merger. }
\label{fig4}
\end{figure*}

To evaluate the precision of the sky maps produced with each method, we consider the area of the 90 percent credible region, as shown in Figure~\ref{fig4}.
The 90 percent credibe region is the area of the smallest credible region containing 90 percent of the posterior probablity.  
It is common for this credible region to be used to set the maximum area that is surveyed in electromagnetic follow up studies (see e.g.,~\cite{Coughlin:2020fwx} and references therein), making this a particularly important quantity. 
For the cases plotted in Figure~\ref{fig4}, the reweighting process generally causes the 90 percent area to increase. For the BBH signals, the 90 percent area blows up and indicates that we cannot recover a precise location when so much information is lost about the signal. For the BNS signals, the 90 percent area increases but not a significant amount. For NSBH signals, the 90 percent area remains the same after the reweighting process.
The increase in the 90 percent area can be understood as accounting for the loss of information due to inpainting and is therefore expected. 

\subsection{Data with a glitch}
\label{sec:glitch}

\begin{figure*}
    \centering
    \includegraphics[width=\textwidth]{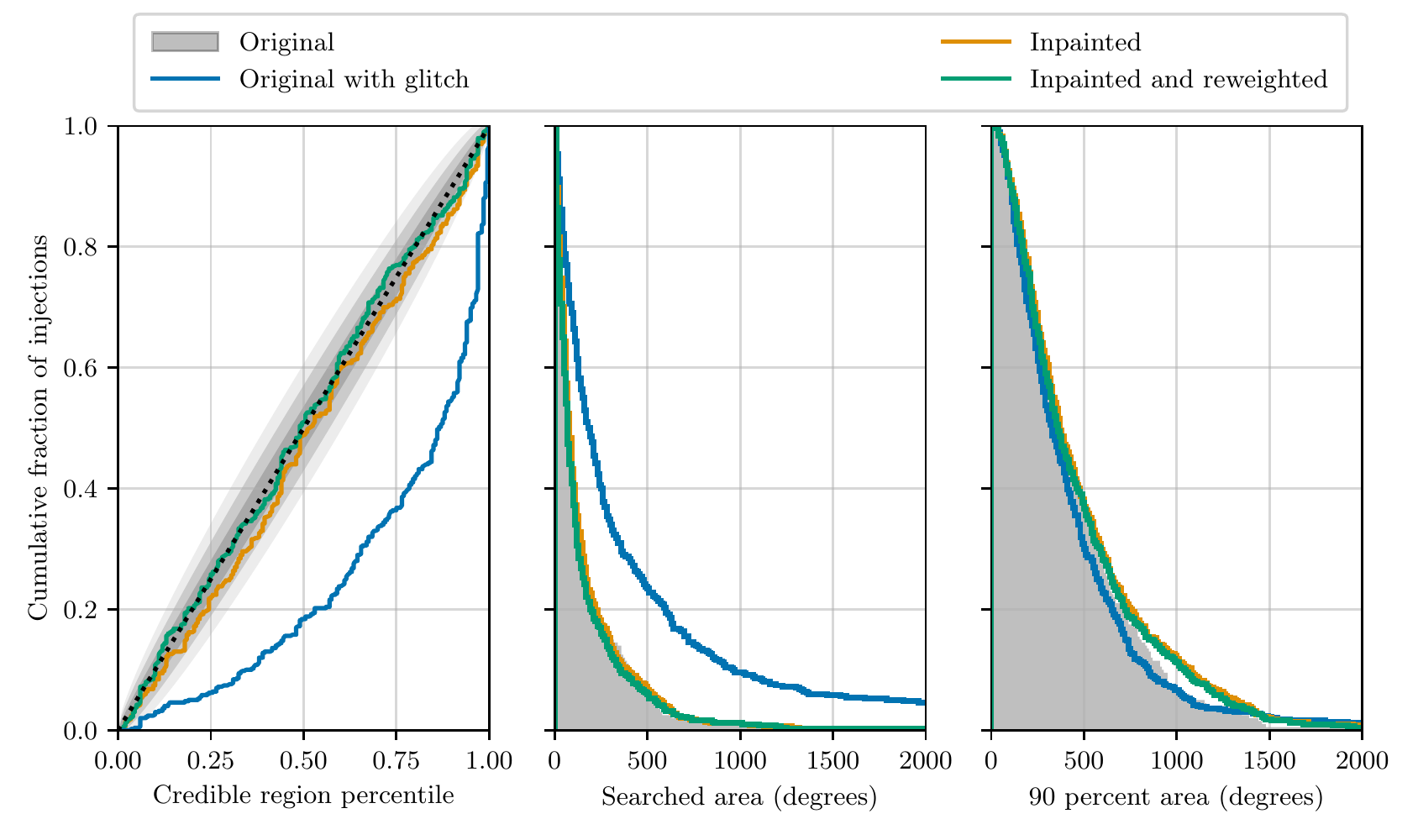}
    \caption{ Distributions of the credible region percentile, searched area, and 90 percent area when a glitch is added to the data. All three panels consider a set of low-mass BBH injections with a simulated glitch added to the data close to the time of merger. The distribution of each statistic without a glitch added is shown as a dotted line with 1-, 2-, and 3-$\sigma$ errors (for the left panel) or a shaded region (for the center and right panels). Also included are the distribution for data with a glitch but with no mitigation (blue), with only inpainting (orange), and with both inpainting and reweighting (green). For all three statistics, we find that the distribution for data with a glitch but no inpainting is skewed with respect to the expectation, showing how this glitch biases the estimated sky localization. We also find that inpainting and reweighting results in a lower searched area and 90 percent area compared to only inpainting.
    }
    \label{fig6}
\end{figure*}

In the previous section, all tests were conducted using colored Gaussian noise. 
Data that is consistent with Gaussian noise would not benefit from the application of this method, meaning that these tests also show less accurate and less precise results after inpainting and reweighting. 
In order to demonstrate how this method can lead to improved estimates of the sky localization of signals, we introduce a simulated glitch on top of our signals and repeat the same studies for one choice of inpainting configuration. 
We simulate a BBH merger with 10 M$_\odot$ component masses and add a sine-Gaussian burst with a frequency of 80 Hz on top of the injected signal.
We then excise the glitch using the same inpainting and reweighting methods. 
We also calculate the same statistics when no glitch is included for comparison purposes. 
The results of this test are shown in Figure~\ref{fig6}.

When including a glitch, we find that inpainting and reweighting is both more precise and accurate than inpainting alone and leads to greatly improved sky localizations compared to no mitigation at all. 
For the P-P panel (left) in Figure~\ref{fig6}, we see that the glitch significant biases the error estimate in BAYESTAR. Inpainting corrects for some of the error, and reweighting leads to an even more accurate estimate than inpainting. 
The searched area panel (center) displays a similar behavior. We see that a glitch biases the accuracy of the skymap in the original data, and it is recovered best by reweighting.
After inpainting and reweighting, the searched area distribution is consistent with the result for no glitch and no mitigation.
The area of the 90 percent credible region (right) shows the inpainted and reweighted histograms are quite close, but larger than the original data with or without the glitch included. 
As previously mentioned, this behavior is expected due to the information loss from inapinting. 

\section{Conclusions}
\label{sec:Conclusions}

As the presence of noise transients may bias parameter estimation for GW signals, 
robust mitigation methods are needed to address any identified data quality issues. 
This is particularly important for low latency applications, where trustworthy information is needed as quick as possible. 
In this work we have presented a solution to this challenge for localizing the source of an identified gravitational-wave signal.
Although it is difficult to know if a specific glitch leads to a bias in the localization of a signal, we have shown that our method can quickly produce sky maps that are known to unbiased in a variety of scenarios. 

The reweighting method discussed in this work is shown to be particularly helpful when significant portions of the signal need to be inpainted due to a glitch
as removing data via the inpainting process (or other similar methods) contributes its own bias. This was a noticeable effect even with a gate of 64 ms (less than a tenth of a second) that we used on the tests shown previously. Other tests we conduced show the same bias for larger gate widths that could be necessary for certain types of observed glitches such as slow-scattering ~\cite{LIGO:2020zwl}. 
We therefore expect PySLIDE to lead to significant improvements in the sky localization in these scenarios as compared to previously utilized methods. 

Although useful in a wide variety of situations, we did identify a number of limitations of this method. 
In cases where the excised data includes the time of merger, we find that the produced sky maps are biased, likely due to some of the assumptions made about the relationship between the SNR and the sky localization accuracy. 
This may be possible to mitigate by accounting for the change in signal bandwidth after inpainting; we reserve this investigation for future works.
We also note that in cases where large portions of the signal are removed by inpainting, there are less benefits of using this method as opposed to simply excluding the impacted detector from the analysis entirely. 
While this method is still applicable, excluding a detector is both more straightforward and guaranteed to not introduce biases in the sky localization. 

In further observing runs we expect the sensitivity of the detectors to increase and to detect more events. 
If the data quality in these runs is similar to past runs, this means that it is more likely there will be instances of glitches overlapping signals from BNS mergers. 
The method presented here can easily handle these types of noise transients, and is particularly useful compared to other methods when long gates are necessary. 
Future work will involve packaging this method into a tool that can be used in low latency, and exploring how we can investigate the source of the previously outlined limitations of the reweighting method.

For upcoming observing runs, it is imperative that we have a way to to mitigate instrumental artifacts in the detector instantaneously. 
Quick and reliable sky localization of GW signals will help facilitate additional multi-messenger detections. 
Our method provides a way to maximize the efficiency of these complex searches and strive towards reliable and useful information even in the presence of noise artifacts.

\section*{Acknowledgements}
The authors thank the other participants and mentors in the LIGO Caltech REU program for productive discussions and support.
We thank Ronaldas Macas for their comments during internal review of this manuscript.
This work is supported by National Science Foundation grant PHY-1852081 as part of the LIGO Caltech REU program. 

This material is based upon work supported by NSF’s LIGO Laboratory 
which is a major facility fully funded by the 
National Science Foundation.
LIGO was constructed by the California Institute of Technology 
and Massachusetts Institute of Technology with funding from 
the National Science Foundation, 
and operates under cooperative agreement PHY-1764464. 
Advanced LIGO was built under award PHY-0823459.
The authors are grateful for computational resources provided by the 
LIGO Laboratory and supported by 
National Science Foundation Grants PHY-0757058 and PHY-0823459.
This work carries LIGO document number P2200195.

\appendix

\section{Complete results}
\label{app:all_results}

For each choices of masses, we completed 9 sets of 200 injections with varying inpainting window sizes and offsets from the time of merger. 
In this appendix, we present results for all of these analyses and discuss the effectiveness of limitations of this method, 
The specific masses, windows, and offsets for these analyses are listed in Table~\ref{tab:config}.

\begin{figure*}
    \centering
    \includegraphics[width=\textwidth]{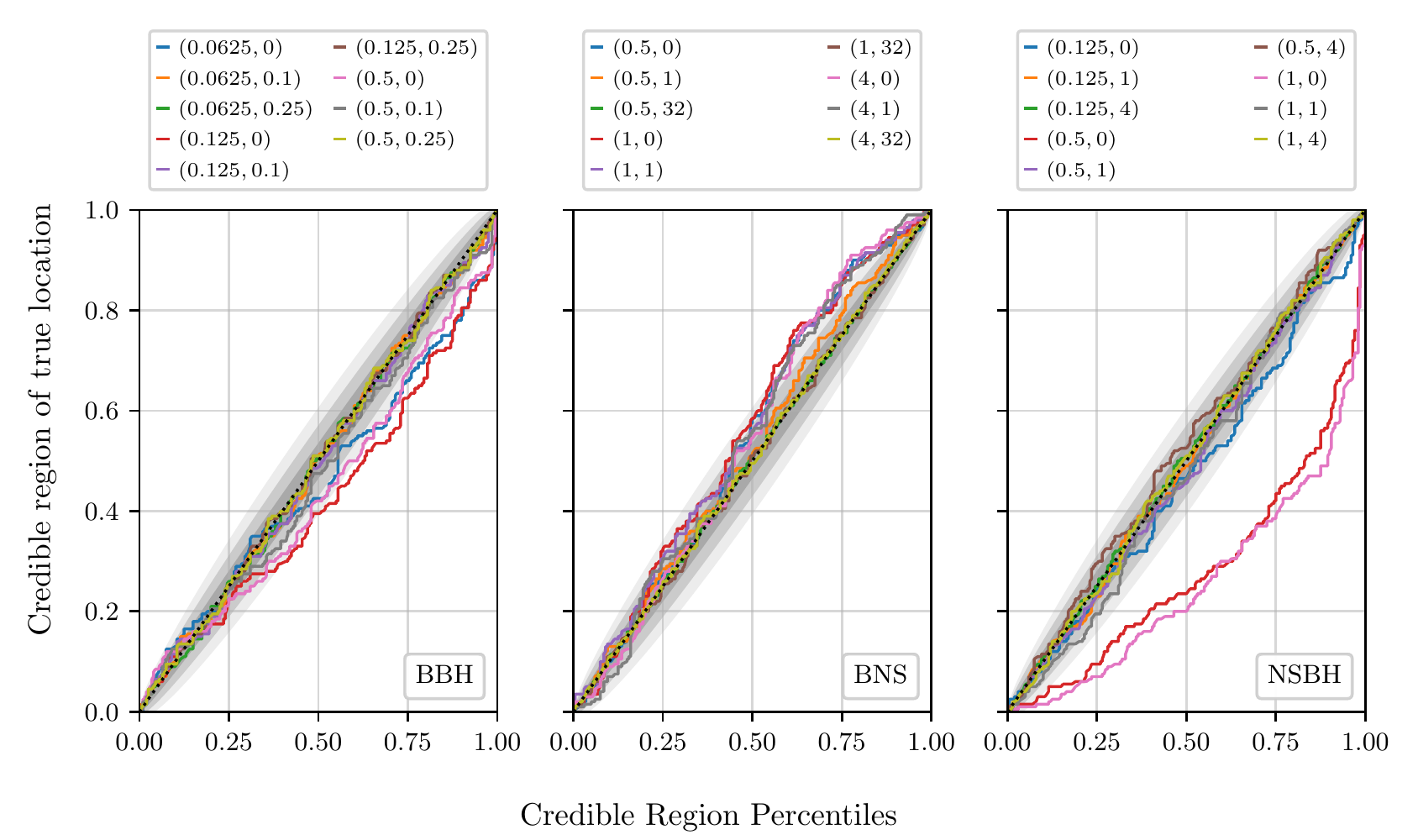}
    \caption{The cumulative fraction of simulations that find the true location of the injected event at each credible level for all scenarios tested in this work, comparable to Figure~\ref{fig2}. Each panel corresponds to simulations with BBH, NSBH, or BNS masses. The legends show the (window size, offset from merger) for each test. All results are normalized with respect to the result measured with no additional data processing. The vast majority of window sizes and offsets show minimal biases, but biases are still observed in cases where the inpainting window directly overlaps the time of merger. The legends show the (window size, offset from merger) for each test.}
    \label{fig:combined_pp}
\end{figure*}

The cumulative fraction of events identified at each credible interval for all tests is shown in Figure~\ref{fig:combined_pp}.
As is done in the body of this paper, we normalize the plotted distribution based on the value with no mitigation to account for any biases inherent to BAYESTAR.
The majority of configurations have results consistent with the expected one-to-one line. However, there are both specific configurations that depart from this line, as well as general trends for each signal class. 
The cases with the largest departures are when the inpainting window directly overlaps the signal's time of merger. 
These differences are especially larger in the NSBH case. 
The BNS results also show a systematic overestimation of the error in cases when the inpainting has a non-trivial impact on the result. 
However, these differences are small compared to the ``merger overlap'' scenario. 

Our results for how the 90 percent credible region is affected by this method are shown in Figure~\ref{fig:combined_search}. 
We find that, in general, we are able to recover the same level of accuracy with our inpainted and reweighted results as the original analysis. 
However, there are notable exceptions for all three signal classes. 
Similar to the previous Figure, these exceptions correspond to cases when large amounts of data are inpainted directly overlapping the signal's time of merger.

\begin{figure*}
    \centering
    \includegraphics[width=\textwidth]{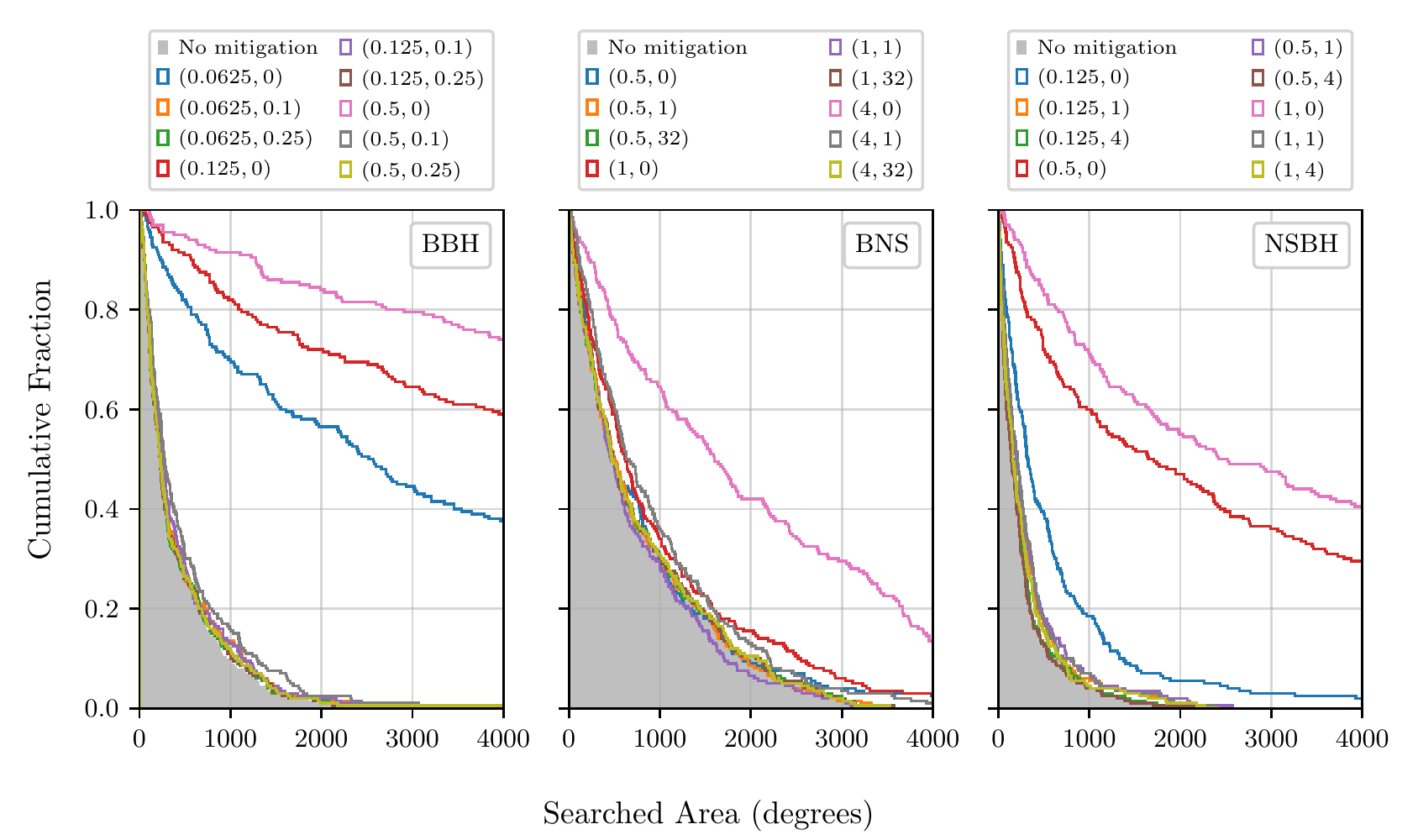}
    \caption{Cumulative fraction of injections with a serarched ares of that size or smaller, comparable to Figure~\ref{fig3}. Each panel corresponds to simulations with BBH, NSBH, or BNS masses. The legends show the (window size, offset from merger) for each test. Ideally, the plotted distributions should match the ``No mitigation'' case. While most distribution follow this expectations, there are a few cases that vastly differ from this expectation. These differences are all from cases where the inpainting window directly overlaps the time of merger.  }
    \label{fig:combined_90}
\end{figure*}

Our results for how the 90 percent credible region is affected by this method are shown in Figure~\ref{fig:combined_90}. 
For this statistic, the results show a continuum from almost perfect agreement with the original analysis to order of magnitude increases in the 90 percent credible area. 
This behavior is expected from this method; when the data is inpainted, the loss of information should increase the error region. 
This increase in area of the 90 percent region is, in general, correctly representing the amount of information loss. 
This figure should be cross-referenced with Figure~\ref{fig:combined_pp} to understand if the reported errors are logically consistent. 

\begin{figure*}
    \centering
    \includegraphics[width=\textwidth]{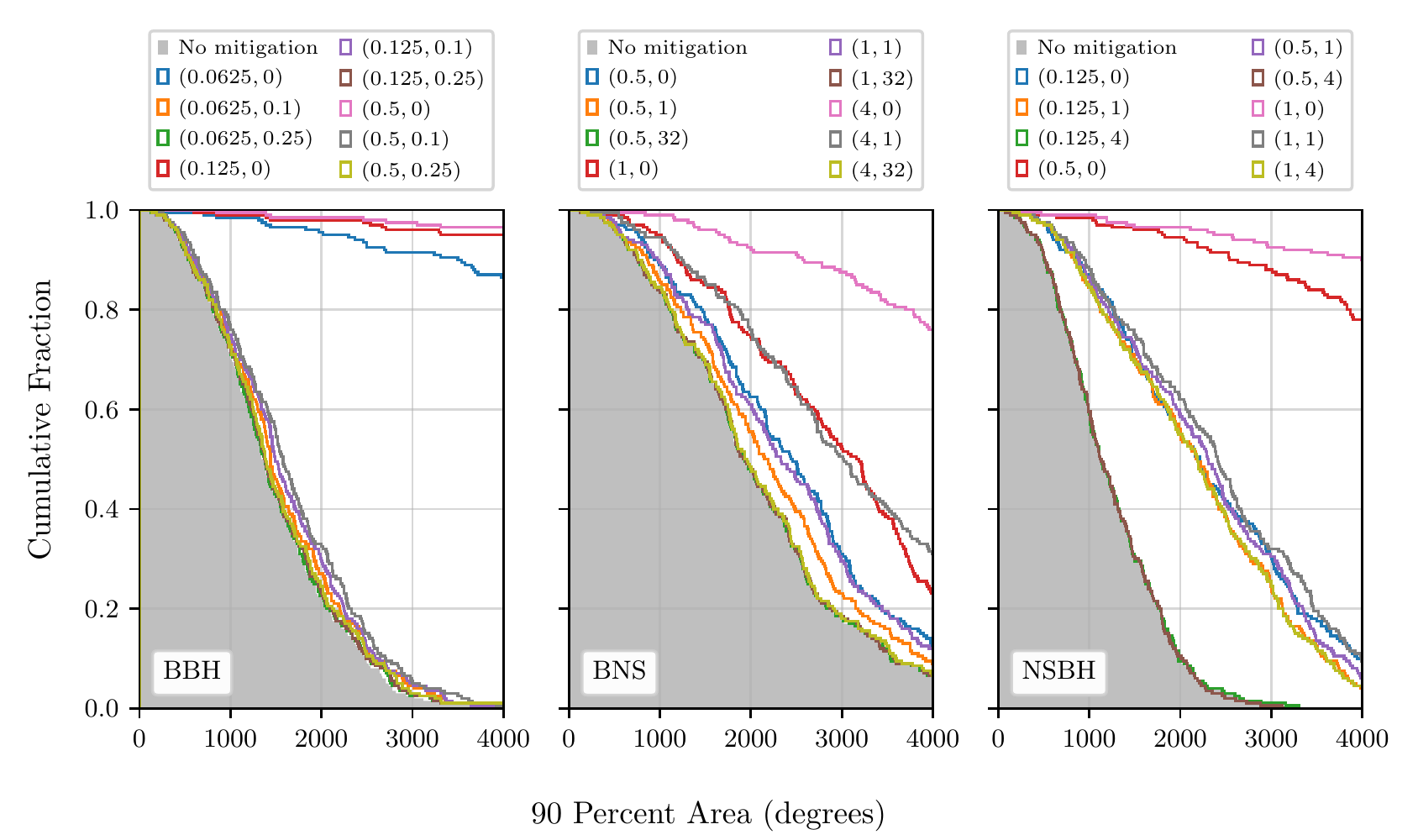}
    \caption{Cumulative fraction of injections with a 90 percent credible area of that size or smaller, comparable to Figure~\ref{fig4}. Each panel corresponds to simulations with BBH, NSBH, or BNS masses. The legends show the (window size, offset from merger) for each test. As inpainting removes data from the analysis, it is expected that all inpainted distributions will skew towards higher values compared to the ``No mitigation'' case. 
    Similar to previous figures, the cases with the larger differences compared to the expectation are those with inpainting widows directly overlapping the time of merger.}
    \label{fig:combined_search}
\end{figure*}

Overall, we find that in cases when the inpainting window does not directly overlap the signal's time of merger, the method presented in this work generates skymaps that are bias free. 
Cases where the inpainting window overlaps the time of merger do show some biases. 
This is likely due to this method assuming that the information contained in each data segment relevant for sky localization is proportional to the SNR of the signal in that time window. 
However, the bandwidth of the signal also plays a subdominant role. 
This assumption is less valid for time periods overlapping the time of merger, where the signal rapidly increases in frequency. 
Despite this limitation, this method may still be beneficial in these cases, as is also known that the presence of a glitch at this time introduces the largest potential biases in the sky localization and other parameters~\cite{Macas:2022afm, powell2018parameter}. 

An additional consideration is if the data is informative after inpainting. 
Although we show that the skymaps produced by this method are logically consistent after removing large portions of the signal, this scenario is similar to simply excluding the effected detector from the analysis. 
Excluding a detector is simpler than this method and should not introduce biases. 
We therefore recommend that this approach be used in cases when the majority of the signal SNR would be removed by inpainting. 

\bibliography{main.bbl}

\begin{thebibliography}{67}%
\makeatletter
\providecommand \@ifxundefined [1]{%
 \@ifx{#1\undefined}
}%
\providecommand \@ifnum [1]{%
 \ifnum #1\expandafter \@firstoftwo
 \else \expandafter \@secondoftwo
 \fi
}%
\providecommand \@ifx [1]{%
 \ifx #1\expandafter \@firstoftwo
 \else \expandafter \@secondoftwo
 \fi
}%
\providecommand \natexlab [1]{#1}%
\providecommand \enquote  [1]{``#1''}%
\providecommand \bibnamefont  [1]{#1}%
\providecommand \bibfnamefont [1]{#1}%
\providecommand \citenamefont [1]{#1}%
\providecommand \href@noop [0]{\@secondoftwo}%
\providecommand \href [0]{\begingroup \@sanitize@url \@href}%
\providecommand \@href[1]{\@@startlink{#1}\@@href}%
\providecommand \@@href[1]{\endgroup#1\@@endlink}%
\providecommand \@sanitize@url [0]{\catcode `\\12\catcode `\$12\catcode
  `\&12\catcode `\#12\catcode `\^12\catcode `\_12\catcode `\%12\relax}%
\providecommand \@@startlink[1]{}%
\providecommand \@@endlink[0]{}%
\providecommand \url  [0]{\begingroup\@sanitize@url \@url }%
\providecommand \@url [1]{\endgroup\@href {#1}{\urlprefix }}%
\providecommand \urlprefix  [0]{URL }%
\providecommand \Eprint [0]{\href }%
\providecommand \doibase [0]{https://doi.org/}%
\providecommand \selectlanguage [0]{\@gobble}%
\providecommand \bibinfo  [0]{\@secondoftwo}%
\providecommand \bibfield  [0]{\@secondoftwo}%
\providecommand \translation [1]{[#1]}%
\providecommand \BibitemOpen [0]{}%
\providecommand \bibitemStop [0]{}%
\providecommand \bibitemNoStop [0]{.\EOS\space}%
\providecommand \EOS [0]{\spacefactor3000\relax}%
\providecommand \BibitemShut  [1]{\csname bibitem#1\endcsname}%
\let\auto@bib@innerbib\@empty
\bibitem [{\citenamefont {Aasi}\ \emph {et~al.}(2015)\citenamefont {Aasi} \emph
  {et~al.}}]{LIGOScientific:2014pky}%
  \BibitemOpen
  \bibfield  {author} {\bibinfo {author} {\bibfnamefont {J.}~\bibnamefont
  {Aasi}} \emph {et~al.} (\bibinfo {collaboration} {LIGO Scientific}),\
  }\bibfield  {title} {\bibinfo {title} {{Advanced LIGO}},\ }\href
  {https://doi.org/10.1088/0264-9381/32/7/074001} {\bibfield  {journal}
  {\bibinfo  {journal} {Class. Quant. Grav.}\ }\textbf {\bibinfo {volume}
  {32}},\ \bibinfo {pages} {074001} (\bibinfo {year} {2015})},\ \Eprint
  {https://arxiv.org/abs/1411.4547} {arXiv:1411.4547 [gr-qc]} \BibitemShut
  {NoStop}%
\bibitem [{\citenamefont {Acernese}\ \emph {et~al.}(2015)\citenamefont
  {Acernese} \emph {et~al.}}]{TheVirgo:2014hva}%
  \BibitemOpen
  \bibfield  {author} {\bibinfo {author} {\bibfnamefont {F.}~\bibnamefont
  {Acernese}} \emph {et~al.} (\bibinfo {collaboration} {VIRGO}),\ }\bibfield
  {title} {\bibinfo {title} {{Advanced Virgo: a second-generation
  interferometric gravitational wave detector}},\ }\href
  {https://doi.org/10.1088/0264-9381/32/2/024001} {\bibfield  {journal}
  {\bibinfo  {journal} {Class. Quantum Grav.}\ }\textbf {\bibinfo {volume}
  {32}},\ \bibinfo {pages} {024001} (\bibinfo {year} {2015})},\ \Eprint
  {https://arxiv.org/abs/1408.3978} {arXiv:1408.3978 [gr-qc]} \BibitemShut
  {NoStop}%
\bibitem [{\citenamefont {Akutsu}\ \emph {et~al.}(2019)\citenamefont {Akutsu}
  \emph {et~al.}}]{KAGRA:2018plz}%
  \BibitemOpen
  \bibfield  {author} {\bibinfo {author} {\bibfnamefont {T.}~\bibnamefont
  {Akutsu}} \emph {et~al.} (\bibinfo {collaboration} {KAGRA}),\ }\bibfield
  {title} {\bibinfo {title} {{KAGRA: 2.5 Generation Interferometric
  Gravitational Wave Detector}},\ }\href
  {https://doi.org/10.1038/s41550-018-0658-y} {\bibfield  {journal} {\bibinfo
  {journal} {Nature Astron.}\ }\textbf {\bibinfo {volume} {3}},\ \bibinfo
  {pages} {35} (\bibinfo {year} {2019})},\ \Eprint
  {https://arxiv.org/abs/1811.08079} {arXiv:1811.08079 [gr-qc]} \BibitemShut
  {NoStop}%
\bibitem [{\citenamefont {Abbott}\ \emph {et~al.}(2019)\citenamefont {Abbott}
  \emph {et~al.}}]{GWTC-1}%
  \BibitemOpen
  \bibfield  {author} {\bibinfo {author} {\bibfnamefont {B.~P.}\ \bibnamefont
  {Abbott}} \emph {et~al.} (\bibinfo {collaboration} {LIGO Scientific,
  Virgo}),\ }\bibfield  {title} {\bibinfo {title} {{GWTC-1: A
  Gravitational-Wave Transient Catalog of Compact Binary Mergers Observed by
  LIGO and Virgo during the First and Second Observing Runs}},\ }\href
  {https://doi.org/10.1103/PhysRevX.9.031040} {\bibfield  {journal} {\bibinfo
  {journal} {Phys. Rev. X}\ ,\ \bibinfo {pages} {031040}} (\bibinfo {year}
  {2019})},\ \Eprint {https://arxiv.org/abs/1811.12907} {arXiv:1811.12907
  [astro-ph.HE]} \BibitemShut {NoStop}%
\bibitem [{\citenamefont {Abbott}\ \emph
  {et~al.}(2021{\natexlab{a}})\citenamefont {Abbott} \emph {et~al.}}]{GWTC-2}%
  \BibitemOpen
  \bibfield  {author} {\bibinfo {author} {\bibfnamefont {R.}~\bibnamefont
  {Abbott}} \emph {et~al.} (\bibinfo {collaboration} {LIGO Scientific,
  Virgo}),\ }\bibfield  {title} {\bibinfo {title} {{GWTC-2: Compact Binary
  Coalescences Observed by LIGO and Virgo During the First Half of the Third
  Observing Run}},\ }\href {https://doi.org/10.1103/PhysRevX.11.021053}
  {\bibfield  {journal} {\bibinfo  {journal} {Phys. Rev. X}\ }\textbf {\bibinfo
  {volume} {11}},\ \bibinfo {pages} {021053} (\bibinfo {year}
  {2021}{\natexlab{a}})},\ \Eprint {https://arxiv.org/abs/2010.14527}
  {arXiv:2010.14527 [gr-qc]} \BibitemShut {NoStop}%
\bibitem [{\citenamefont {Abbott}\ \emph
  {et~al.}(2021{\natexlab{b}})\citenamefont {Abbott} \emph {et~al.}}]{GWTC-3}%
  \BibitemOpen
  \bibfield  {author} {\bibinfo {author} {\bibfnamefont {R.}~\bibnamefont
  {Abbott}} \emph {et~al.} (\bibinfo {collaboration} {LIGO Scientific, VIRGO,
  KAGRA}),\ }\bibfield  {title} {\bibinfo {title} {{GWTC-3: Compact Binary
  Coalescences Observed by LIGO and Virgo During the Second Part of the Third
  Observing Run}},\ }\href@noop {} {\  (\bibinfo {year}
  {2021}{\natexlab{b}})},\ \Eprint {https://arxiv.org/abs/2111.03606}
  {arXiv:2111.03606 [gr-qc]} \BibitemShut {NoStop}%
\bibitem [{\citenamefont {Abbott}\ \emph
  {et~al.}(2021{\natexlab{c}})\citenamefont {Abbott} \emph
  {et~al.}}]{GWTC-2.1}%
  \BibitemOpen
  \bibfield  {author} {\bibinfo {author} {\bibfnamefont {R.}~\bibnamefont
  {Abbott}} \emph {et~al.} (\bibinfo {collaboration} {LIGO Scientific,
  VIRGO}),\ }\bibfield  {title} {\bibinfo {title} {{GWTC-2.1: Deep Extended
  Catalog of Compact Binary Coalescences Observed by LIGO and Virgo During the
  First Half of the Third Observing Run}},\ }\href@noop {} {\  (\bibinfo {year}
  {2021}{\natexlab{c}})},\ \Eprint {https://arxiv.org/abs/2108.01045}
  {arXiv:2108.01045 [gr-qc]} \BibitemShut {NoStop}%
\bibitem [{\citenamefont {Abbott}\ \emph {et~al.}(2017)\citenamefont {Abbott}
  \emph {et~al.}}]{PhysRevLett.119.161101}%
  \BibitemOpen
  \bibfield  {author} {\bibinfo {author} {\bibfnamefont {B.~P.}\ \bibnamefont
  {Abbott}} \emph {et~al.} (\bibinfo {collaboration} {LIGO Scientific
  Collaboration and Virgo Collaboration}),\ }\bibfield  {title} {\bibinfo
  {title} {Gw170817: Observation of gravitational waves from a binary neutron
  star inspiral},\ }\href {https://doi.org/10.1103/PhysRevLett.119.161101}
  {\bibfield  {journal} {\bibinfo  {journal} {Phys. Rev. Lett.}\ }\textbf
  {\bibinfo {volume} {119}},\ \bibinfo {pages} {161101} (\bibinfo {year}
  {2017})}\BibitemShut {NoStop}%
\bibitem [{\citenamefont {{Abbott}}\ \emph {et~al.}(2020)\citenamefont
  {{Abbott}} \emph {et~al.}}]{2020ApJ...892L...3A}%
  \BibitemOpen
  \bibfield  {author} {\bibinfo {author} {\bibfnamefont {B.~P.}\ \bibnamefont
  {{Abbott}}} \emph {et~al.},\ }\bibfield  {title} {\bibinfo {title}
  {{GW190425: Observation of a Compact Binary Coalescence with Total Mass
  {\ensuremath{\sim}} 3.4 M$_{{\ensuremath{\odot}}}$}},\ }\href
  {https://doi.org/10.3847/2041-8213/ab75f5} {\bibfield  {journal} {\bibinfo
  {journal} {\apjl}\ }\textbf {\bibinfo {volume} {892}},\ \bibinfo {eid} {L3}
  (\bibinfo {year} {2020})},\ \Eprint {https://arxiv.org/abs/2001.01761}
  {arXiv:2001.01761 [astro-ph.HE]} \BibitemShut {NoStop}%
\bibitem [{\citenamefont {{Abbott}}\ \emph {et~al.}(2021)\citenamefont
  {{Abbott}} \emph {et~al.}}]{2021ApJ...915L...5A}%
  \BibitemOpen
  \bibfield  {author} {\bibinfo {author} {\bibfnamefont {R.}~\bibnamefont
  {{Abbott}}} \emph {et~al.},\ }\bibfield  {title} {\bibinfo {title}
  {{Observation of Gravitational Waves from Two Neutron Star-Black Hole
  Coalescences}},\ }\href {https://doi.org/10.3847/2041-8213/ac082e} {\bibfield
   {journal} {\bibinfo  {journal} {\apjl}\ }\textbf {\bibinfo {volume} {915}},\
  \bibinfo {eid} {L5} (\bibinfo {year} {2021})},\ \Eprint
  {https://arxiv.org/abs/2106.15163} {arXiv:2106.15163 [astro-ph.HE]}
  \BibitemShut {NoStop}%
\bibitem [{\citenamefont {{Abbott}}\ \emph {et~al.}(2017)\citenamefont
  {{Abbott}} \emph {et~al.}}]{2017ApJ...848L..13A}%
  \BibitemOpen
  \bibfield  {author} {\bibinfo {author} {\bibfnamefont {B.~P.}\ \bibnamefont
  {{Abbott}}} \emph {et~al.},\ }\bibfield  {title} {\bibinfo {title}
  {{Gravitational Waves and Gamma-Rays from a Binary Neutron Star Merger:
  GW170817 and GRB 170817A}},\ }\href
  {https://doi.org/10.3847/2041-8213/aa920c} {\bibfield  {journal} {\bibinfo
  {journal} {Astrophysical Journal Letters}\ }\textbf {\bibinfo {volume}
  {848}},\ \bibinfo {eid} {L13} (\bibinfo {year} {2017})},\ \Eprint
  {https://arxiv.org/abs/1710.05834} {arXiv:1710.05834 [astro-ph.HE]}
  \BibitemShut {NoStop}%
\bibitem [{\citenamefont {{Goldstein}}\ \emph {et~al.}(2017)\citenamefont
  {{Goldstein}} \emph {et~al.}}]{2017ApJ...848L..14G}%
  \BibitemOpen
  \bibfield  {author} {\bibinfo {author} {\bibfnamefont {A.}~\bibnamefont
  {{Goldstein}}} \emph {et~al.},\ }\bibfield  {title} {\bibinfo {title} {{An
  Ordinary Short Gamma-Ray Burst with Extraordinary Implications: Fermi-GBM
  Detection of GRB 170817A}},\ }\href
  {https://doi.org/10.3847/2041-8213/aa8f41} {\bibfield  {journal} {\bibinfo
  {journal} {Astrophysical Journal Letters}\ }\textbf {\bibinfo {volume}
  {848}},\ \bibinfo {eid} {L14} (\bibinfo {year} {2017})},\ \Eprint
  {https://arxiv.org/abs/1710.05446} {arXiv:1710.05446 [astro-ph.HE]}
  \BibitemShut {NoStop}%
\bibitem [{\citenamefont {Hallinan}\ \emph {et~al.}(2017)\citenamefont
  {Hallinan}, \citenamefont {Corsi}, \citenamefont {Mooley}, \citenamefont
  {Hotokezaka}, \citenamefont {Nakar}, \citenamefont {Kasliwal}, \citenamefont
  {Kaplan}, \citenamefont {Frail}, \citenamefont {Myers}, \citenamefont
  {Murphy}, \citenamefont {De}, \citenamefont {Dobie}, \citenamefont {Allison},
  \citenamefont {Bannister}, \citenamefont {Bhalerao}, \citenamefont {Chandra},
  \citenamefont {Clarke}, \citenamefont {Giacintucci}, \citenamefont {Ho},
  \citenamefont {Horesh}, \citenamefont {Kassim}, \citenamefont {Kulkarni},
  \citenamefont {Lenc}, \citenamefont {Lockman}, \citenamefont {Lynch},
  \citenamefont {Nichols}, \citenamefont {Nissanke}, \citenamefont
  {Palliyaguru}, \citenamefont {Peters}, \citenamefont {Piran}, \citenamefont
  {Rana}, \citenamefont {Sadler},\ and\ \citenamefont
  {Singer}}]{doi:10.1126/science.aap9855}%
  \BibitemOpen
  \bibfield  {author} {\bibinfo {author} {\bibfnamefont {G.}~\bibnamefont
  {Hallinan}}, \bibinfo {author} {\bibfnamefont {A.}~\bibnamefont {Corsi}},
  \bibinfo {author} {\bibfnamefont {K.~P.}\ \bibnamefont {Mooley}}, \bibinfo
  {author} {\bibfnamefont {K.}~\bibnamefont {Hotokezaka}}, \bibinfo {author}
  {\bibfnamefont {E.}~\bibnamefont {Nakar}}, \bibinfo {author} {\bibfnamefont
  {M.~M.}\ \bibnamefont {Kasliwal}}, \bibinfo {author} {\bibfnamefont {D.~L.}\
  \bibnamefont {Kaplan}}, \bibinfo {author} {\bibfnamefont {D.~A.}\
  \bibnamefont {Frail}}, \bibinfo {author} {\bibfnamefont {S.~T.}\ \bibnamefont
  {Myers}}, \bibinfo {author} {\bibfnamefont {T.}~\bibnamefont {Murphy}},
  \bibinfo {author} {\bibfnamefont {K.}~\bibnamefont {De}}, \bibinfo {author}
  {\bibfnamefont {D.}~\bibnamefont {Dobie}}, \bibinfo {author} {\bibfnamefont
  {J.~R.}\ \bibnamefont {Allison}}, \bibinfo {author} {\bibfnamefont {K.~W.}\
  \bibnamefont {Bannister}}, \bibinfo {author} {\bibfnamefont {V.}~\bibnamefont
  {Bhalerao}}, \bibinfo {author} {\bibfnamefont {P.}~\bibnamefont {Chandra}},
  \bibinfo {author} {\bibfnamefont {T.~E.}\ \bibnamefont {Clarke}}, \bibinfo
  {author} {\bibfnamefont {S.}~\bibnamefont {Giacintucci}}, \bibinfo {author}
  {\bibfnamefont {A.~Y.~Q.}\ \bibnamefont {Ho}}, \bibinfo {author}
  {\bibfnamefont {A.}~\bibnamefont {Horesh}}, \bibinfo {author} {\bibfnamefont
  {N.~E.}\ \bibnamefont {Kassim}}, \bibinfo {author} {\bibfnamefont {S.~R.}\
  \bibnamefont {Kulkarni}}, \bibinfo {author} {\bibfnamefont {E.}~\bibnamefont
  {Lenc}}, \bibinfo {author} {\bibfnamefont {F.~J.}\ \bibnamefont {Lockman}},
  \bibinfo {author} {\bibfnamefont {C.}~\bibnamefont {Lynch}}, \bibinfo
  {author} {\bibfnamefont {D.}~\bibnamefont {Nichols}}, \bibinfo {author}
  {\bibfnamefont {S.}~\bibnamefont {Nissanke}}, \bibinfo {author}
  {\bibfnamefont {N.}~\bibnamefont {Palliyaguru}}, \bibinfo {author}
  {\bibfnamefont {W.~M.}\ \bibnamefont {Peters}}, \bibinfo {author}
  {\bibfnamefont {T.}~\bibnamefont {Piran}}, \bibinfo {author} {\bibfnamefont
  {J.}~\bibnamefont {Rana}}, \bibinfo {author} {\bibfnamefont {E.~M.}\
  \bibnamefont {Sadler}},\ and\ \bibinfo {author} {\bibfnamefont {L.~P.}\
  \bibnamefont {Singer}},\ }\bibfield  {title} {\bibinfo {title} {A radio
  counterpart to a neutron star merger},\ }\href
  {https://doi.org/10.1126/science.aap9855} {\bibfield  {journal} {\bibinfo
  {journal} {Science}\ }\textbf {\bibinfo {volume} {358}},\ \bibinfo {pages}
  {1579} (\bibinfo {year} {2017})},\ \Eprint
  {https://arxiv.org/abs/https://www.science.org/doi/pdf/10.1126/science.aap9855}
  {https://www.science.org/doi/pdf/10.1126/science.aap9855} \BibitemShut
  {NoStop}%
\bibitem [{\citenamefont {{Alexander}}\ \emph {et~al.}(2017)\citenamefont
  {{Alexander}} \emph {et~al.}}]{2017ApJ...848L..21A}%
  \BibitemOpen
  \bibfield  {author} {\bibinfo {author} {\bibfnamefont {K.~D.}\ \bibnamefont
  {{Alexander}}} \emph {et~al.},\ }\bibfield  {title} {\bibinfo {title} {{The
  Electromagnetic Counterpart of the Binary Neutron Star Merger LIGO/Virgo
  GW170817. VI. Radio Constraints on a Relativistic Jet and Predictions for
  Late-time Emission from the Kilonova Ejecta}},\ }\href
  {https://doi.org/10.3847/2041-8213/aa905d} {\bibfield  {journal} {\bibinfo
  {journal} {Astrophysical Journal Letters}\ }\textbf {\bibinfo {volume}
  {848}},\ \bibinfo {eid} {L21} (\bibinfo {year} {2017})},\ \Eprint
  {https://arxiv.org/abs/1710.05457} {arXiv:1710.05457 [astro-ph.HE]}
  \BibitemShut {NoStop}%
\bibitem [{\citenamefont {{Troja}}\ \emph {et~al.}(2017)\citenamefont {{Troja}}
  \emph {et~al.}}]{2017Natur.551...71T}%
  \BibitemOpen
  \bibfield  {author} {\bibinfo {author} {\bibfnamefont {E.}~\bibnamefont
  {{Troja}}} \emph {et~al.},\ }\bibfield  {title} {\bibinfo {title} {{The X-ray
  counterpart to the gravitational-wave event GW170817}},\ }\href
  {https://doi.org/10.1038/nature24290} {\bibfield  {journal} {\bibinfo
  {journal} {\nat}\ }\textbf {\bibinfo {volume} {551}},\ \bibinfo {pages} {71}
  (\bibinfo {year} {2017})},\ \Eprint {https://arxiv.org/abs/1710.05433}
  {arXiv:1710.05433 [astro-ph.HE]} \BibitemShut {NoStop}%
\bibitem [{\citenamefont {{Lyman}}\ \emph {et~al.}(2018)\citenamefont {{Lyman}}
  \emph {et~al.}}]{2018NatAs...2..751L}%
  \BibitemOpen
  \bibfield  {author} {\bibinfo {author} {\bibfnamefont {J.~D.}\ \bibnamefont
  {{Lyman}}} \emph {et~al.},\ }\bibfield  {title} {\bibinfo {title} {{The
  optical afterglow of the short gamma-ray burst associated with GW170817}},\
  }\href {https://doi.org/10.1038/s41550-018-0511-3} {\bibfield  {journal}
  {\bibinfo  {journal} {Nature Astronomy}\ }\textbf {\bibinfo {volume} {2}},\
  \bibinfo {pages} {751} (\bibinfo {year} {2018})},\ \Eprint
  {https://arxiv.org/abs/1801.02669} {arXiv:1801.02669 [astro-ph.HE]}
  \BibitemShut {NoStop}%
\bibitem [{\citenamefont {Balasubramanian}\ \emph {et~al.}(2021)\citenamefont
  {Balasubramanian}, \citenamefont {Corsi}, \citenamefont {Mooley},
  \citenamefont {Brightman}, \citenamefont {Hallinan}, \citenamefont
  {Hotokezaka}, \citenamefont {Kaplan}, \citenamefont {Lazzati},\ and\
  \citenamefont {Murphy}}]{Balasubramanian:2021kny}%
  \BibitemOpen
  \bibfield  {author} {\bibinfo {author} {\bibfnamefont {A.}~\bibnamefont
  {Balasubramanian}}, \bibinfo {author} {\bibfnamefont {A.}~\bibnamefont
  {Corsi}}, \bibinfo {author} {\bibfnamefont {K.~P.}\ \bibnamefont {Mooley}},
  \bibinfo {author} {\bibfnamefont {M.}~\bibnamefont {Brightman}}, \bibinfo
  {author} {\bibfnamefont {G.}~\bibnamefont {Hallinan}}, \bibinfo {author}
  {\bibfnamefont {K.}~\bibnamefont {Hotokezaka}}, \bibinfo {author}
  {\bibfnamefont {D.~L.}\ \bibnamefont {Kaplan}}, \bibinfo {author}
  {\bibfnamefont {D.}~\bibnamefont {Lazzati}},\ and\ \bibinfo {author}
  {\bibfnamefont {E.~J.}\ \bibnamefont {Murphy}},\ }\bibfield  {title}
  {\bibinfo {title} {{Continued Radio Observations of GW170817 3.5 yr
  Post-merger}},\ }\href {https://doi.org/10.3847/2041-8213/abfd38} {\bibfield
  {journal} {\bibinfo  {journal} {Astrophys. J. Lett.}\ }\textbf {\bibinfo
  {volume} {914}},\ \bibinfo {pages} {L20} (\bibinfo {year} {2021})},\ \Eprint
  {https://arxiv.org/abs/2103.04821} {arXiv:2103.04821 [astro-ph.HE]}
  \BibitemShut {NoStop}%
\bibitem [{\citenamefont {Metzger}(2020)}]{Metzger:2019zeh}%
  \BibitemOpen
  \bibfield  {author} {\bibinfo {author} {\bibfnamefont {B.~D.}\ \bibnamefont
  {Metzger}},\ }\bibfield  {title} {\bibinfo {title} {{Kilonovae}},\ }\href
  {https://doi.org/10.1007/s41114-019-0024-0} {\bibfield  {journal} {\bibinfo
  {journal} {Living Rev. Rel.}\ }\textbf {\bibinfo {volume} {23}},\ \bibinfo
  {pages} {1} (\bibinfo {year} {2020})},\ \Eprint
  {https://arxiv.org/abs/1910.01617} {arXiv:1910.01617 [astro-ph.HE]}
  \BibitemShut {NoStop}%
\bibitem [{\citenamefont {{Chornock}}\ \emph {et~al.}(2017)\citenamefont
  {{Chornock}} \emph {et~al.}}]{2017ApJ...848L..19C}%
  \BibitemOpen
  \bibfield  {author} {\bibinfo {author} {\bibfnamefont {R.}~\bibnamefont
  {{Chornock}}} \emph {et~al.},\ }\bibfield  {title} {\bibinfo {title} {{The
  Electromagnetic Counterpart of the Binary Neutron Star Merger LIGO/Virgo
  GW170817. IV. Detection of Near-infrared Signatures of r-process
  Nucleosynthesis with Gemini-South}},\ }\href
  {https://doi.org/10.3847/2041-8213/aa905c} {\bibfield  {journal} {\bibinfo
  {journal} {Astrophysical Journal Letters}\ }\textbf {\bibinfo {volume}
  {848}},\ \bibinfo {eid} {L19} (\bibinfo {year} {2017})},\ \Eprint
  {https://arxiv.org/abs/1710.05454} {arXiv:1710.05454 [astro-ph.HE]}
  \BibitemShut {NoStop}%
\bibitem [{\citenamefont {{Kasliwal}}\ \emph {et~al.}(2017)\citenamefont
  {{Kasliwal}} \emph {et~al.}}]{2017Sci...358.1559K}%
  \BibitemOpen
  \bibfield  {author} {\bibinfo {author} {\bibfnamefont {M.~M.}\ \bibnamefont
  {{Kasliwal}}} \emph {et~al.},\ }\bibfield  {title} {\bibinfo {title}
  {{Illuminating gravitational waves: A concordant picture of photons from a
  neutron star merger}},\ }\href {https://doi.org/10.1126/science.aap9455}
  {\bibfield  {journal} {\bibinfo  {journal} {Science}\ }\textbf {\bibinfo
  {volume} {358}},\ \bibinfo {pages} {1559} (\bibinfo {year} {2017})},\ \Eprint
  {https://arxiv.org/abs/1710.05436} {arXiv:1710.05436 [astro-ph.HE]}
  \BibitemShut {NoStop}%
\bibitem [{\citenamefont {Abbott}\ \emph
  {et~al.}(2017{\natexlab{a}})\citenamefont {Abbott} \emph
  {et~al.}}]{LIGOScientific:2017ync}%
  \BibitemOpen
  \bibfield  {author} {\bibinfo {author} {\bibfnamefont {B.~P.}\ \bibnamefont
  {Abbott}} \emph {et~al.} (\bibinfo {collaboration} {LIGO Scientific, Virgo,
  Fermi GBM, INTEGRAL, IceCube, AstroSat Cadmium Zinc Telluride Imager Team,
  IPN, Insight-Hxmt, ANTARES, Swift, AGILE Team, 1M2H Team, Dark Energy Camera
  GW-EM, DES, DLT40, GRAWITA, Fermi-LAT, ATCA, ASKAP, Las Cumbres Observatory
  Group, OzGrav, DWF (Deeper Wider Faster Program), AST3, CAASTRO, VINROUGE,
  MASTER, J-GEM, GROWTH, JAGWAR, CaltechNRAO, TTU-NRAO, NuSTAR, Pan-STARRS,
  MAXI Team, TZAC Consortium, KU, Nordic Optical Telescope, ePESSTO, GROND,
  Texas Tech University, SALT Group, TOROS, BOOTES, MWA, CALET, IKI-GW
  Follow-up, H.E.S.S., LOFAR, LWA, HAWC, Pierre Auger, ALMA, Euro VLBI Team, Pi
  of Sky, Chandra Team at McGill University, DFN, ATLAS Telescopes, High Time
  Resolution Universe Survey, RIMAS, RATIR, SKA South Africa/MeerKAT}),\
  }\bibfield  {title} {\bibinfo {title} {{Multi-messenger Observations of a
  Binary Neutron Star Merger}},\ }\href
  {https://doi.org/10.3847/2041-8213/aa91c9} {\bibfield  {journal} {\bibinfo
  {journal} {Astrophys. J. Lett.}\ }\textbf {\bibinfo {volume} {848}},\
  \bibinfo {pages} {L12} (\bibinfo {year} {2017}{\natexlab{a}})},\ \Eprint
  {https://arxiv.org/abs/1710.05833} {arXiv:1710.05833 [astro-ph.HE]}
  \BibitemShut {NoStop}%
\bibitem [{\citenamefont {Abbott}\ \emph
  {et~al.}(2021{\natexlab{a}})\citenamefont {Abbott} \emph
  {et~al.}}]{PhysRevD.103.122002}%
  \BibitemOpen
  \bibfield  {author} {\bibinfo {author} {\bibfnamefont {R.}~\bibnamefont
  {Abbott}} \emph {et~al.} (\bibinfo {collaboration} {LIGO Scientific
  Collaboration and Virgo Collaboration}),\ }\bibfield  {title} {\bibinfo
  {title} {Tests of general relativity with binary black holes from the second
  ligo-virgo gravitational-wave transient catalog},\ }\href
  {https://doi.org/10.1103/PhysRevD.103.122002} {\bibfield  {journal} {\bibinfo
   {journal} {Phys. Rev. D}\ }\textbf {\bibinfo {volume} {103}},\ \bibinfo
  {pages} {122002} (\bibinfo {year} {2021}{\natexlab{a}})}\BibitemShut
  {NoStop}%
\bibitem [{\citenamefont {Abbott}\ \emph {et~al.}(2018)\citenamefont {Abbott}
  \emph {et~al.}}]{LIGOScientific:2018cki}%
  \BibitemOpen
  \bibfield  {author} {\bibinfo {author} {\bibfnamefont {B.~P.}\ \bibnamefont
  {Abbott}} \emph {et~al.} (\bibinfo {collaboration} {LIGO Scientific,
  Virgo}),\ }\bibfield  {title} {\bibinfo {title} {{GW170817: Measurements of
  neutron star radii and equation of state}},\ }\href
  {https://doi.org/10.1103/PhysRevLett.121.161101} {\bibfield  {journal}
  {\bibinfo  {journal} {Phys. Rev. Lett.}\ }\textbf {\bibinfo {volume} {121}},\
  \bibinfo {pages} {161101} (\bibinfo {year} {2018})},\ \Eprint
  {https://arxiv.org/abs/1805.11581} {arXiv:1805.11581 [gr-qc]} \BibitemShut
  {NoStop}%
\bibitem [{\citenamefont {{The LIGO Scientific Collaboration}}\ \emph
  {et~al.}(2021)\citenamefont {{The LIGO Scientific Collaboration}} \emph
  {et~al.}}]{2021arXiv211103634T}%
  \BibitemOpen
  \bibfield  {author} {\bibinfo {author} {\bibnamefont {{The LIGO Scientific
  Collaboration}}} \emph {et~al.},\ }\bibfield  {title} {\bibinfo {title} {{The
  population of merging compact binaries inferred using gravitational waves
  through GWTC-3}},\ }\href@noop {} {\bibfield  {journal} {\bibinfo  {journal}
  {arXiv e-prints}\ ,\ \bibinfo {eid} {arXiv:2111.03634}} (\bibinfo {year}
  {2021})},\ \Eprint {https://arxiv.org/abs/2111.03634} {arXiv:2111.03634
  [astro-ph.HE]} \BibitemShut {NoStop}%
\bibitem [{\citenamefont {Abbott}\ \emph {et~al.}(2020)\citenamefont {Abbott},
  \citenamefont {Abbott}, \citenamefont {Abraham}, \citenamefont {Acernese},
  \citenamefont {Ackley}, \citenamefont {Adams}, \citenamefont {Adhikari},
  \citenamefont {Adya}, \citenamefont {Affeldt}, \citenamefont {Agathos} \emph
  {et~al.}}]{abbott2020gw190521}%
  \BibitemOpen
  \bibfield  {author} {\bibinfo {author} {\bibfnamefont {R.}~\bibnamefont
  {Abbott}}, \bibinfo {author} {\bibfnamefont {T.}~\bibnamefont {Abbott}},
  \bibinfo {author} {\bibfnamefont {S.}~\bibnamefont {Abraham}}, \bibinfo
  {author} {\bibfnamefont {F.}~\bibnamefont {Acernese}}, \bibinfo {author}
  {\bibfnamefont {K.}~\bibnamefont {Ackley}}, \bibinfo {author} {\bibfnamefont
  {C.}~\bibnamefont {Adams}}, \bibinfo {author} {\bibfnamefont
  {R.}~\bibnamefont {Adhikari}}, \bibinfo {author} {\bibfnamefont
  {V.}~\bibnamefont {Adya}}, \bibinfo {author} {\bibfnamefont {C.}~\bibnamefont
  {Affeldt}}, \bibinfo {author} {\bibfnamefont {M.}~\bibnamefont {Agathos}},
  \emph {et~al.},\ }\bibfield  {title} {\bibinfo {title} {{Gw190521: A binary
  black hole merger with a total mass of 150 M$_{{\ensuremath{\odot}}}$}},\
  }\href@noop {} {\bibfield  {journal} {\bibinfo  {journal} {Physical review
  letters}\ }\textbf {\bibinfo {volume} {125}},\ \bibinfo {pages} {101102}
  (\bibinfo {year} {2020})}\BibitemShut {NoStop}%
\bibitem [{\citenamefont {{Abbott}}\ \emph {et~al.}(2020)\citenamefont
  {{Abbott}} \emph {et~al.}}]{2020ApJ...896L..44A}%
  \BibitemOpen
  \bibfield  {author} {\bibinfo {author} {\bibfnamefont {R.}~\bibnamefont
  {{Abbott}}} \emph {et~al.},\ }\bibfield  {title} {\bibinfo {title}
  {{GW190814: Gravitational Waves from the Coalescence of a 23 Solar Mass Black
  Hole with a 2.6 Solar Mass Compact Object}},\ }\href
  {https://doi.org/10.3847/2041-8213/ab960f} {\bibfield  {journal} {\bibinfo
  {journal} {ApJ}\ }\textbf {\bibinfo {volume} {896}},\ \bibinfo {eid} {L44}
  (\bibinfo {year} {2020})},\ \Eprint {https://arxiv.org/abs/2006.12611}
  {arXiv:2006.12611 [astro-ph.HE]} \BibitemShut {NoStop}%
\bibitem [{\citenamefont {{Fairhurst}}(2011)}]{2011CQGra..28j5021F}%
  \BibitemOpen
  \bibfield  {author} {\bibinfo {author} {\bibfnamefont {S.}~\bibnamefont
  {{Fairhurst}}},\ }\bibfield  {title} {\bibinfo {title} {{Source localization
  with an advanced gravitational wave detector network}},\ }\href
  {https://doi.org/10.1088/0264-9381/28/10/105021} {\bibfield  {journal}
  {\bibinfo  {journal} {Classical and Quantum Gravity}\ }\textbf {\bibinfo
  {volume} {28}},\ \bibinfo {eid} {105021} (\bibinfo {year} {2011})},\ \Eprint
  {https://arxiv.org/abs/1010.6192} {arXiv:1010.6192 [gr-qc]} \BibitemShut
  {NoStop}%
\bibitem [{\citenamefont {{Singer}}\ \emph {et~al.}(2016)\citenamefont
  {{Singer}} \emph {et~al.}}]{2016ApJ...829L..15S}%
  \BibitemOpen
  \bibfield  {author} {\bibinfo {author} {\bibfnamefont {L.~P.}\ \bibnamefont
  {{Singer}}} \emph {et~al.},\ }\bibfield  {title} {\bibinfo {title} {{Going
  the Distance: Mapping Host Galaxies of LIGO and Virgo Sources in Three
  Dimensions Using Local Cosmography and Targeted Follow-up}},\ }\href
  {https://doi.org/10.3847/2041-8205/829/1/L15} {\bibfield  {journal} {\bibinfo
   {journal} {Astrophysical Journal Letters}\ }\textbf {\bibinfo {volume}
  {829}},\ \bibinfo {eid} {L15} (\bibinfo {year} {2016})},\ \Eprint
  {https://arxiv.org/abs/1603.07333} {arXiv:1603.07333 [astro-ph.HE]}
  \BibitemShut {NoStop}%
\bibitem [{\citenamefont {{Abbott}}\ \emph {et~al.}(2018)\citenamefont
  {{Abbott}} \emph {et~al.}}]{2018LRR....21....3A}%
  \BibitemOpen
  \bibfield  {author} {\bibinfo {author} {\bibfnamefont {B.~P.}\ \bibnamefont
  {{Abbott}}} \emph {et~al.},\ }\bibfield  {title} {\bibinfo {title}
  {{Prospects for observing and localizing gravitational-wave transients with
  Advanced LIGO, Advanced Virgo and KAGRA}},\ }\href
  {https://doi.org/10.1007/s41114-018-0012-9} {\bibfield  {journal} {\bibinfo
  {journal} {Living Reviews in Relativity}\ }\textbf {\bibinfo {volume} {21}},\
  \bibinfo {eid} {3} (\bibinfo {year} {2018})},\ \Eprint
  {https://arxiv.org/abs/1304.0670} {arXiv:1304.0670 [gr-qc]} \BibitemShut
  {NoStop}%
\bibitem [{\citenamefont {Coughlin}\ \emph {et~al.}(2018)\citenamefont
  {Coughlin} \emph {et~al.}}]{Coughlin:2018lta}%
  \BibitemOpen
  \bibfield  {author} {\bibinfo {author} {\bibfnamefont {M.~W.}\ \bibnamefont
  {Coughlin}} \emph {et~al.},\ }\bibfield  {title} {\bibinfo {title}
  {{Optimizing searches for electromagnetic counterparts of gravitational wave
  triggers}},\ }\href {https://doi.org/10.1093/mnras/sty1066} {\bibfield
  {journal} {\bibinfo  {journal} {Mon. Not. Roy. Astron. Soc.}\ }\textbf
  {\bibinfo {volume} {478}},\ \bibinfo {pages} {692} (\bibinfo {year}
  {2018})},\ \Eprint {https://arxiv.org/abs/1803.02255} {arXiv:1803.02255
  [astro-ph.IM]} \BibitemShut {NoStop}%
\bibitem [{\citenamefont {Coughlin}\ \emph {et~al.}(2019)\citenamefont
  {Coughlin} \emph {et~al.}}]{Coughlin:2019qkn}%
  \BibitemOpen
  \bibfield  {author} {\bibinfo {author} {\bibfnamefont {M.~W.}\ \bibnamefont
  {Coughlin}} \emph {et~al.},\ }\bibfield  {title} {\bibinfo {title}
  {{Optimizing multitelescope observations of gravitational-wave
  counterparts}},\ }\href {https://doi.org/10.1093/mnras/stz2485} {\bibfield
  {journal} {\bibinfo  {journal} {Mon. Not. Roy. Astron. Soc.}\ }\textbf
  {\bibinfo {volume} {489}},\ \bibinfo {pages} {5775} (\bibinfo {year}
  {2019})},\ \Eprint {https://arxiv.org/abs/1909.01244} {arXiv:1909.01244
  [astro-ph.IM]} \BibitemShut {NoStop}%
\bibitem [{\citenamefont {Wen}\ and\ \citenamefont
  {Chen}(2010)}]{PhysRevD.81.082001}%
  \BibitemOpen
  \bibfield  {author} {\bibinfo {author} {\bibfnamefont {L.}~\bibnamefont
  {Wen}}\ and\ \bibinfo {author} {\bibfnamefont {Y.}~\bibnamefont {Chen}},\
  }\bibfield  {title} {\bibinfo {title} {Geometrical expression for the angular
  resolution of a network of gravitational-wave detectors},\ }\href
  {https://doi.org/10.1103/PhysRevD.81.082001} {\bibfield  {journal} {\bibinfo
  {journal} {Phys. Rev. D}\ }\textbf {\bibinfo {volume} {81}},\ \bibinfo
  {pages} {082001} (\bibinfo {year} {2010})}\BibitemShut {NoStop}%
\bibitem [{\citenamefont {{Nissanke}}\ \emph {et~al.}(2013)\citenamefont
  {{Nissanke}}, \citenamefont {{Kasliwal}},\ and\ \citenamefont
  {{Georgieva}}}]{2013ApJ...767..124N}%
  \BibitemOpen
  \bibfield  {author} {\bibinfo {author} {\bibfnamefont {S.}~\bibnamefont
  {{Nissanke}}}, \bibinfo {author} {\bibfnamefont {M.}~\bibnamefont
  {{Kasliwal}}},\ and\ \bibinfo {author} {\bibfnamefont {A.}~\bibnamefont
  {{Georgieva}}},\ }\bibfield  {title} {\bibinfo {title} {{Identifying Elusive
  Electromagnetic Counterparts to Gravitational Wave Mergers: An End-to-end
  Simulation}},\ }\href {https://doi.org/10.1088/0004-637X/767/2/124}
  {\bibfield  {journal} {\bibinfo  {journal} {\apj}\ }\textbf {\bibinfo
  {volume} {767}},\ \bibinfo {eid} {124} (\bibinfo {year} {2013})},\ \Eprint
  {https://arxiv.org/abs/1210.6362} {arXiv:1210.6362 [astro-ph.HE]}
  \BibitemShut {NoStop}%
\bibitem [{\citenamefont {Sidery}\ \emph {et~al.}(2014)\citenamefont {Sidery},
  \citenamefont {Aylott}, \citenamefont {Christensen}, \citenamefont {Farr},
  \citenamefont {Farr}, \citenamefont {Feroz}, \citenamefont {Gair},
  \citenamefont {Grover}, \citenamefont {Graff}, \citenamefont {Hanna},
  \citenamefont {Kalogera}, \citenamefont {Mandel}, \citenamefont
  {O'Shaughnessy}, \citenamefont {Pitkin}, \citenamefont {Price}, \citenamefont
  {Raymond}, \citenamefont {R\"over}, \citenamefont {Singer}, \citenamefont
  {van~der Sluys}, \citenamefont {Smith}, \citenamefont {Vecchio},
  \citenamefont {Veitch},\ and\ \citenamefont {Vitale}}]{PhysRevD.89.084060}%
  \BibitemOpen
  \bibfield  {author} {\bibinfo {author} {\bibfnamefont {T.}~\bibnamefont
  {Sidery}}, \bibinfo {author} {\bibfnamefont {B.}~\bibnamefont {Aylott}},
  \bibinfo {author} {\bibfnamefont {N.}~\bibnamefont {Christensen}}, \bibinfo
  {author} {\bibfnamefont {B.}~\bibnamefont {Farr}}, \bibinfo {author}
  {\bibfnamefont {W.}~\bibnamefont {Farr}}, \bibinfo {author} {\bibfnamefont
  {F.}~\bibnamefont {Feroz}}, \bibinfo {author} {\bibfnamefont
  {J.}~\bibnamefont {Gair}}, \bibinfo {author} {\bibfnamefont {K.}~\bibnamefont
  {Grover}}, \bibinfo {author} {\bibfnamefont {P.}~\bibnamefont {Graff}},
  \bibinfo {author} {\bibfnamefont {C.}~\bibnamefont {Hanna}}, \bibinfo
  {author} {\bibfnamefont {V.}~\bibnamefont {Kalogera}}, \bibinfo {author}
  {\bibfnamefont {I.}~\bibnamefont {Mandel}}, \bibinfo {author} {\bibfnamefont
  {R.}~\bibnamefont {O'Shaughnessy}}, \bibinfo {author} {\bibfnamefont
  {M.}~\bibnamefont {Pitkin}}, \bibinfo {author} {\bibfnamefont
  {L.}~\bibnamefont {Price}}, \bibinfo {author} {\bibfnamefont
  {V.}~\bibnamefont {Raymond}}, \bibinfo {author} {\bibfnamefont
  {C.}~\bibnamefont {R\"over}}, \bibinfo {author} {\bibfnamefont
  {L.}~\bibnamefont {Singer}}, \bibinfo {author} {\bibfnamefont
  {M.}~\bibnamefont {van~der Sluys}}, \bibinfo {author} {\bibfnamefont
  {R.~J.~E.}\ \bibnamefont {Smith}}, \bibinfo {author} {\bibfnamefont
  {A.}~\bibnamefont {Vecchio}}, \bibinfo {author} {\bibfnamefont
  {J.}~\bibnamefont {Veitch}},\ and\ \bibinfo {author} {\bibfnamefont
  {S.}~\bibnamefont {Vitale}},\ }\bibfield  {title} {\bibinfo {title}
  {Reconstructing the sky location of gravitational-wave detected compact
  binary systems: Methodology for testing and comparison},\ }\href
  {https://doi.org/10.1103/PhysRevD.89.084060} {\bibfield  {journal} {\bibinfo
  {journal} {Phys. Rev. D}\ }\textbf {\bibinfo {volume} {89}},\ \bibinfo
  {pages} {084060} (\bibinfo {year} {2014})}\BibitemShut {NoStop}%
\bibitem [{\citenamefont {Macas}\ \emph {et~al.}(2022)\citenamefont {Macas},
  \citenamefont {Pooley}, \citenamefont {Nuttall}, \citenamefont {Davis},
  \citenamefont {Dyer}, \citenamefont {Lecoeuche}, \citenamefont {Lyman},
  \citenamefont {McIver},\ and\ \citenamefont {Rink}}]{Macas:2022afm}%
  \BibitemOpen
  \bibfield  {author} {\bibinfo {author} {\bibfnamefont {R.}~\bibnamefont
  {Macas}}, \bibinfo {author} {\bibfnamefont {J.}~\bibnamefont {Pooley}},
  \bibinfo {author} {\bibfnamefont {L.~K.}\ \bibnamefont {Nuttall}}, \bibinfo
  {author} {\bibfnamefont {D.}~\bibnamefont {Davis}}, \bibinfo {author}
  {\bibfnamefont {M.~J.}\ \bibnamefont {Dyer}}, \bibinfo {author}
  {\bibfnamefont {Y.}~\bibnamefont {Lecoeuche}}, \bibinfo {author}
  {\bibfnamefont {J.~D.}\ \bibnamefont {Lyman}}, \bibinfo {author}
  {\bibfnamefont {J.}~\bibnamefont {McIver}},\ and\ \bibinfo {author}
  {\bibfnamefont {K.}~\bibnamefont {Rink}},\ }\bibfield  {title} {\bibinfo
  {title} {{Impact of noise transients on low latency gravitational-wave event
  localization}},\ }\href {https://doi.org/10.1103/PhysRevD.105.103021}
  {\bibfield  {journal} {\bibinfo  {journal} {Phys. Rev. D}\ }\textbf {\bibinfo
  {volume} {105}},\ \bibinfo {pages} {103021} (\bibinfo {year} {2022})},\
  \Eprint {https://arxiv.org/abs/2202.00344} {arXiv:2202.00344 [astro-ph.HE]}
  \BibitemShut {NoStop}%
\bibitem [{\citenamefont {Powell}(2018)}]{powell2018parameter}%
  \BibitemOpen
  \bibfield  {author} {\bibinfo {author} {\bibfnamefont {J.}~\bibnamefont
  {Powell}},\ }\bibfield  {title} {\bibinfo {title} {Parameter estimation and
  model selection of gravitational wave signals contaminated by transient
  detector noise glitches},\ }\href@noop {} {\bibfield  {journal} {\bibinfo
  {journal} {Classical and Quantum Gravity}\ }\textbf {\bibinfo {volume}
  {35}},\ \bibinfo {pages} {155017} (\bibinfo {year} {2018})}\BibitemShut
  {NoStop}%
\bibitem [{\citenamefont {Cabero}\ \emph {et~al.}(2020)\citenamefont {Cabero},
  \citenamefont {Mahabal},\ and\ \citenamefont {McIver}}]{Cabero:2020eik}%
  \BibitemOpen
  \bibfield  {author} {\bibinfo {author} {\bibfnamefont {M.}~\bibnamefont
  {Cabero}}, \bibinfo {author} {\bibfnamefont {A.}~\bibnamefont {Mahabal}},\
  and\ \bibinfo {author} {\bibfnamefont {J.}~\bibnamefont {McIver}},\
  }\bibfield  {title} {\bibinfo {title} {{GWSkyNet: a real-time classifier for
  public gravitational-wave candidates}},\ }\href
  {https://doi.org/10.3847/2041-8213/abc5b5} {\bibfield  {journal} {\bibinfo
  {journal} {Astrophys. J. Lett.}\ }\textbf {\bibinfo {volume} {904}},\
  \bibinfo {pages} {L9} (\bibinfo {year} {2020})},\ \Eprint
  {https://arxiv.org/abs/2010.11829} {arXiv:2010.11829 [gr-qc]} \BibitemShut
  {NoStop}%
\bibitem [{\citenamefont {Vitale}\ \emph {et~al.}(2017)\citenamefont {Vitale},
  \citenamefont {Essick}, \citenamefont {Katsavounidis}, \citenamefont
  {Klimenko},\ and\ \citenamefont {Vedovato}}]{Vitale:2016jlv}%
  \BibitemOpen
  \bibfield  {author} {\bibinfo {author} {\bibfnamefont {S.}~\bibnamefont
  {Vitale}}, \bibinfo {author} {\bibfnamefont {R.}~\bibnamefont {Essick}},
  \bibinfo {author} {\bibfnamefont {E.}~\bibnamefont {Katsavounidis}}, \bibinfo
  {author} {\bibfnamefont {S.}~\bibnamefont {Klimenko}},\ and\ \bibinfo
  {author} {\bibfnamefont {G.}~\bibnamefont {Vedovato}},\ }\bibfield  {title}
  {\bibinfo {title} {{On similarity of binary black hole gravitational-wave
  skymaps: to observe or to wait?}},\ }\href
  {https://doi.org/10.1093/mnrasl/slw239} {\bibfield  {journal} {\bibinfo
  {journal} {Mon. Not. Roy. Astron. Soc.}\ }\textbf {\bibinfo {volume} {466}},\
  \bibinfo {pages} {L78} (\bibinfo {year} {2017})},\ \Eprint
  {https://arxiv.org/abs/1611.02438} {arXiv:1611.02438 [gr-qc]} \BibitemShut
  {NoStop}%
\bibitem [{\citenamefont {Davis}\ \emph
  {et~al.}(2021{\natexlab{a}})\citenamefont {Davis} \emph
  {et~al.}}]{LIGO:2021ppb}%
  \BibitemOpen
  \bibfield  {author} {\bibinfo {author} {\bibfnamefont {D.}~\bibnamefont
  {Davis}} \emph {et~al.} (\bibinfo {collaboration} {LIGO}),\ }\bibfield
  {title} {\bibinfo {title} {{LIGO detector characterization in the second and
  third observing runs}},\ }\href {https://doi.org/10.1088/1361-6382/abfd85}
  {\bibfield  {journal} {\bibinfo  {journal} {Class. Quant. Grav.}\ }\textbf
  {\bibinfo {volume} {38}},\ \bibinfo {pages} {135014} (\bibinfo {year}
  {2021}{\natexlab{a}})},\ \Eprint {https://arxiv.org/abs/2101.11673}
  {arXiv:2101.11673 [astro-ph.IM]} \BibitemShut {NoStop}%
\bibitem [{\citenamefont {Aasi}\ \emph {et~al.}(2012)\citenamefont {Aasi},
  \citenamefont {Abadie}, \citenamefont {Abbott}, \citenamefont {Abbott},
  \citenamefont {Abbott}, \citenamefont {Abernathy}, \citenamefont {Accadia},
  \citenamefont {Acernese}, \citenamefont {Adams}, \citenamefont {Adams} \emph
  {et~al.}}]{aasi2012characterization}%
  \BibitemOpen
  \bibfield  {author} {\bibinfo {author} {\bibfnamefont {J.}~\bibnamefont
  {Aasi}}, \bibinfo {author} {\bibfnamefont {J.}~\bibnamefont {Abadie}},
  \bibinfo {author} {\bibfnamefont {B.}~\bibnamefont {Abbott}}, \bibinfo
  {author} {\bibfnamefont {R.}~\bibnamefont {Abbott}}, \bibinfo {author}
  {\bibfnamefont {T.}~\bibnamefont {Abbott}}, \bibinfo {author} {\bibfnamefont
  {M.}~\bibnamefont {Abernathy}}, \bibinfo {author} {\bibfnamefont
  {T.}~\bibnamefont {Accadia}}, \bibinfo {author} {\bibfnamefont
  {F.}~\bibnamefont {Acernese}}, \bibinfo {author} {\bibfnamefont
  {C.}~\bibnamefont {Adams}}, \bibinfo {author} {\bibfnamefont
  {T.}~\bibnamefont {Adams}}, \emph {et~al.},\ }\bibfield  {title} {\bibinfo
  {title} {The characterization of virgo data and its impact on
  gravitational-wave searches},\ }\href@noop {} {\bibfield  {journal} {\bibinfo
   {journal} {Classical and Quantum Gravity}\ }\textbf {\bibinfo {volume}
  {29}},\ \bibinfo {pages} {155002} (\bibinfo {year} {2012})}\BibitemShut
  {NoStop}%
\bibitem [{\citenamefont {Akutsu}\ \emph {et~al.}(2021)\citenamefont {Akutsu}
  \emph {et~al.}}]{KAGRA:2020agh}%
  \BibitemOpen
  \bibfield  {author} {\bibinfo {author} {\bibfnamefont {T.}~\bibnamefont
  {Akutsu}} \emph {et~al.} (\bibinfo {collaboration} {KAGRA}),\ }\bibfield
  {title} {\bibinfo {title} {{Overview of KAGRA: Calibration, detector
  characterization, physical environmental monitors, and the geophysics
  interferometer}},\ }\href {https://doi.org/10.1093/ptep/ptab018} {\bibfield
  {journal} {\bibinfo  {journal} {PTEP}\ }\textbf {\bibinfo {volume} {2021}},\
  \bibinfo {pages} {05A102} (\bibinfo {year} {2021})},\ \Eprint
  {https://arxiv.org/abs/2009.09305} {arXiv:2009.09305 [gr-qc]} \BibitemShut
  {NoStop}%
\bibitem [{\citenamefont {Nguyen}\ \emph {et~al.}(2021)\citenamefont {Nguyen},
  \citenamefont {Schofield}, \citenamefont {Effler}, \citenamefont {Austin},
  \citenamefont {Adya}, \citenamefont {Ball}, \citenamefont {Banagiri},
  \citenamefont {Banowetz}, \citenamefont {Billman}, \citenamefont {Blair}
  \emph {et~al.}}]{nguyen2021environmental}%
  \BibitemOpen
  \bibfield  {author} {\bibinfo {author} {\bibfnamefont {P.}~\bibnamefont
  {Nguyen}}, \bibinfo {author} {\bibfnamefont {R.}~\bibnamefont {Schofield}},
  \bibinfo {author} {\bibfnamefont {A.}~\bibnamefont {Effler}}, \bibinfo
  {author} {\bibfnamefont {C.}~\bibnamefont {Austin}}, \bibinfo {author}
  {\bibfnamefont {V.}~\bibnamefont {Adya}}, \bibinfo {author} {\bibfnamefont
  {M.}~\bibnamefont {Ball}}, \bibinfo {author} {\bibfnamefont {S.}~\bibnamefont
  {Banagiri}}, \bibinfo {author} {\bibfnamefont {K.}~\bibnamefont {Banowetz}},
  \bibinfo {author} {\bibfnamefont {C.}~\bibnamefont {Billman}}, \bibinfo
  {author} {\bibfnamefont {C.}~\bibnamefont {Blair}}, \emph {et~al.},\
  }\bibfield  {title} {\bibinfo {title} {Environmental noise in advanced ligo
  detectors},\ }\href@noop {} {\bibfield  {journal} {\bibinfo  {journal}
  {Classical and Quantum Gravity}\ }\textbf {\bibinfo {volume} {38}},\ \bibinfo
  {pages} {145001} (\bibinfo {year} {2021})}\BibitemShut {NoStop}%
\bibitem [{\citenamefont {{Nuttall}}(2018)}]{2018RSPTA.37670286N}%
  \BibitemOpen
  \bibfield  {author} {\bibinfo {author} {\bibfnamefont {L.~K.}\ \bibnamefont
  {{Nuttall}}},\ }\bibfield  {title} {\bibinfo {title} {{Characterizing
  transient noise in the LIGO detectors}},\ }\href
  {https://doi.org/10.1098/rsta.2017.0286} {\bibfield  {journal} {\bibinfo
  {journal} {Philosophical Transactions of the Royal Society of London Series
  A}\ }\textbf {\bibinfo {volume} {376}},\ \bibinfo {eid} {20170286} (\bibinfo
  {year} {2018})},\ \Eprint {https://arxiv.org/abs/1804.07592}
  {arXiv:1804.07592 [astro-ph.IM]} \BibitemShut {NoStop}%
\bibitem [{\citenamefont {Davis}\ \emph
  {et~al.}(2021{\natexlab{b}})\citenamefont {Davis}, \citenamefont {Areeda},
  \citenamefont {Berger}, \citenamefont {Bruntz}, \citenamefont {Effler},
  \citenamefont {Essick}, \citenamefont {Fisher}, \citenamefont {Godwin},
  \citenamefont {Goetz}, \citenamefont {Helmling-Cornell} \emph
  {et~al.}}]{davis2021ligo}%
  \BibitemOpen
  \bibfield  {author} {\bibinfo {author} {\bibfnamefont {D.}~\bibnamefont
  {Davis}}, \bibinfo {author} {\bibfnamefont {J.~S.}\ \bibnamefont {Areeda}},
  \bibinfo {author} {\bibfnamefont {B.~K.}\ \bibnamefont {Berger}}, \bibinfo
  {author} {\bibfnamefont {R.}~\bibnamefont {Bruntz}}, \bibinfo {author}
  {\bibfnamefont {A.}~\bibnamefont {Effler}}, \bibinfo {author} {\bibfnamefont
  {R.}~\bibnamefont {Essick}}, \bibinfo {author} {\bibfnamefont
  {R.}~\bibnamefont {Fisher}}, \bibinfo {author} {\bibfnamefont
  {P.}~\bibnamefont {Godwin}}, \bibinfo {author} {\bibfnamefont
  {E.}~\bibnamefont {Goetz}}, \bibinfo {author} {\bibfnamefont
  {A.}~\bibnamefont {Helmling-Cornell}}, \emph {et~al.},\ }\bibfield  {title}
  {\bibinfo {title} {Ligo detector characterization in the second and third
  observing runs},\ }\href@noop {} {\bibfield  {journal} {\bibinfo  {journal}
  {Classical and Quantum Gravity}\ }\textbf {\bibinfo {volume} {38}},\ \bibinfo
  {pages} {135014} (\bibinfo {year} {2021}{\natexlab{b}})}\BibitemShut
  {NoStop}%
\bibitem [{\citenamefont {{Berger}}(2018)}]{2018JPhCS.957a2004B}%
  \BibitemOpen
  \bibfield  {author} {\bibinfo {author} {\bibfnamefont {B.~K.}\ \bibnamefont
  {{Berger}}},\ }\bibfield  {title} {\bibinfo {title} {{Identification and
  mitigation of Advanced LIGO noise sources}},\ }in\ \href
  {https://doi.org/10.1088/1742-6596/957/1/012004} {\emph {\bibinfo {booktitle}
  {Journal of Physics Conference Series}}},\ \bibinfo {series} {Journal of
  Physics Conference Series}, Vol.\ \bibinfo {volume} {957}\ (\bibinfo {year}
  {2018})\ p.\ \bibinfo {pages} {012004}\BibitemShut {NoStop}%
\bibitem [{\citenamefont {Pankow}\ \emph {et~al.}(2018)\citenamefont {Pankow}
  \emph {et~al.}}]{Pankow:2018qpo}%
  \BibitemOpen
  \bibfield  {author} {\bibinfo {author} {\bibfnamefont {C.}~\bibnamefont
  {Pankow}} \emph {et~al.},\ }\bibfield  {title} {\bibinfo {title} {{Mitigation
  of the instrumental noise transient in gravitational-wave data surrounding
  GW170817}},\ }\href {https://doi.org/10.1103/PhysRevD.98.084016} {\bibfield
  {journal} {\bibinfo  {journal} {Phys. Rev. D}\ }\textbf {\bibinfo {volume}
  {98}},\ \bibinfo {pages} {084016} (\bibinfo {year} {2018})},\ \Eprint
  {https://arxiv.org/abs/1808.03619} {arXiv:1808.03619 [gr-qc]} \BibitemShut
  {NoStop}%
\bibitem [{\citenamefont {Harris}(1978)}]{harris1978use}%
  \BibitemOpen
  \bibfield  {author} {\bibinfo {author} {\bibfnamefont {F.~J.}\ \bibnamefont
  {Harris}},\ }\bibfield  {title} {\bibinfo {title} {On the use of windows for
  harmonic analysis with the discrete fourier transform},\ }\href@noop {}
  {\bibfield  {journal} {\bibinfo  {journal} {Proceedings of the IEEE}\
  }\textbf {\bibinfo {volume} {66}},\ \bibinfo {pages} {51} (\bibinfo {year}
  {1978})}\BibitemShut {NoStop}%
\bibitem [{\citenamefont {Cornish}\ \emph
  {et~al.}(2021{\natexlab{a}})\citenamefont {Cornish}, \citenamefont
  {Littenberg}, \citenamefont {B\'ecsy}, \citenamefont {Chatziioannou},
  \citenamefont {Clark}, \citenamefont {Ghonge},\ and\ \citenamefont
  {Millhouse}}]{Cornish:2020dwh}%
  \BibitemOpen
  \bibfield  {author} {\bibinfo {author} {\bibfnamefont {N.~J.}\ \bibnamefont
  {Cornish}}, \bibinfo {author} {\bibfnamefont {T.~B.}\ \bibnamefont
  {Littenberg}}, \bibinfo {author} {\bibfnamefont {B.}~\bibnamefont {B\'ecsy}},
  \bibinfo {author} {\bibfnamefont {K.}~\bibnamefont {Chatziioannou}}, \bibinfo
  {author} {\bibfnamefont {J.~A.}\ \bibnamefont {Clark}}, \bibinfo {author}
  {\bibfnamefont {S.}~\bibnamefont {Ghonge}},\ and\ \bibinfo {author}
  {\bibfnamefont {M.}~\bibnamefont {Millhouse}},\ }\bibfield  {title} {\bibinfo
  {title} {{BayesWave analysis pipeline in the era of gravitational wave
  observations}},\ }\href {https://doi.org/10.1103/PhysRevD.103.044006}
  {\bibfield  {journal} {\bibinfo  {journal} {Phys. Rev. D}\ }\textbf {\bibinfo
  {volume} {103}},\ \bibinfo {pages} {044006} (\bibinfo {year}
  {2021}{\natexlab{a}})},\ \Eprint {https://arxiv.org/abs/2011.09494}
  {arXiv:2011.09494 [gr-qc]} \BibitemShut {NoStop}%
\bibitem [{\citenamefont {Cornish}\ and\ \citenamefont
  {Littenberg}(2015)}]{cornish2015bayeswave}%
  \BibitemOpen
  \bibfield  {author} {\bibinfo {author} {\bibfnamefont {N.~J.}\ \bibnamefont
  {Cornish}}\ and\ \bibinfo {author} {\bibfnamefont {T.~B.}\ \bibnamefont
  {Littenberg}},\ }\bibfield  {title} {\bibinfo {title} {Bayeswave: Bayesian
  inference for gravitational wave bursts and instrument glitches},\
  }\href@noop {} {\bibfield  {journal} {\bibinfo  {journal} {Classical and
  Quantum Gravity}\ }\textbf {\bibinfo {volume} {32}},\ \bibinfo {pages}
  {135012} (\bibinfo {year} {2015})}\BibitemShut {NoStop}%
\bibitem [{\citenamefont {Cornish}\ \emph
  {et~al.}(2021{\natexlab{b}})\citenamefont {Cornish}, \citenamefont
  {Littenberg}, \citenamefont {B{\'e}csy}, \citenamefont {Chatziioannou},
  \citenamefont {Clark}, \citenamefont {Ghonge},\ and\ \citenamefont
  {Millhouse}}]{cornish2021bayeswave}%
  \BibitemOpen
  \bibfield  {author} {\bibinfo {author} {\bibfnamefont {N.~J.}\ \bibnamefont
  {Cornish}}, \bibinfo {author} {\bibfnamefont {T.~B.}\ \bibnamefont
  {Littenberg}}, \bibinfo {author} {\bibfnamefont {B.}~\bibnamefont
  {B{\'e}csy}}, \bibinfo {author} {\bibfnamefont {K.}~\bibnamefont
  {Chatziioannou}}, \bibinfo {author} {\bibfnamefont {J.~A.}\ \bibnamefont
  {Clark}}, \bibinfo {author} {\bibfnamefont {S.}~\bibnamefont {Ghonge}},\ and\
  \bibinfo {author} {\bibfnamefont {M.}~\bibnamefont {Millhouse}},\ }\bibfield
  {title} {\bibinfo {title} {Bayeswave analysis pipeline in the era of
  gravitational wave observations},\ }\href@noop {} {\bibfield  {journal}
  {\bibinfo  {journal} {Physical Review D}\ }\textbf {\bibinfo {volume}
  {103}},\ \bibinfo {pages} {044006} (\bibinfo {year}
  {2021}{\natexlab{b}})}\BibitemShut {NoStop}%
\bibitem [{\citenamefont {Mogushi}\ \emph {et~al.}(2021)\citenamefont
  {Mogushi}, \citenamefont {Quitzow-James}, \citenamefont {Cavagli\`a},
  \citenamefont {Kulkarni},\ and\ \citenamefont {Hayes}}]{Mogushi:2021cpw}%
  \BibitemOpen
  \bibfield  {author} {\bibinfo {author} {\bibfnamefont {K.}~\bibnamefont
  {Mogushi}}, \bibinfo {author} {\bibfnamefont {R.}~\bibnamefont
  {Quitzow-James}}, \bibinfo {author} {\bibfnamefont {M.}~\bibnamefont
  {Cavagli\`a}}, \bibinfo {author} {\bibfnamefont {S.}~\bibnamefont
  {Kulkarni}},\ and\ \bibinfo {author} {\bibfnamefont {F.}~\bibnamefont
  {Hayes}},\ }\bibfield  {title} {\bibinfo {title} {{NNETFIX: an artificial
  neural network-based denoising engine for gravitational-wave signals}},\
  }\href {https://doi.org/10.1088/2632-2153/abea69} {\bibfield  {journal}
  {\bibinfo  {journal} {Mach. Learn. Sci. Tech.}\ }\textbf {\bibinfo {volume}
  {2}},\ \bibinfo {pages} {035018} (\bibinfo {year} {2021})},\ \Eprint
  {https://arxiv.org/abs/2101.04712} {arXiv:2101.04712 [gr-qc]} \BibitemShut
  {NoStop}%
\bibitem [{\citenamefont {Cuoco}\ \emph {et~al.}(2020)\citenamefont {Cuoco},
  \citenamefont {Powell}, \citenamefont {Cavagli{\`a}}, \citenamefont {Ackley},
  \citenamefont {Bejger}, \citenamefont {Chatterjee}, \citenamefont {Coughlin},
  \citenamefont {Coughlin}, \citenamefont {Easter}, \citenamefont {Essick}
  \emph {et~al.}}]{cuoco2020enhancing}%
  \BibitemOpen
  \bibfield  {author} {\bibinfo {author} {\bibfnamefont {E.}~\bibnamefont
  {Cuoco}}, \bibinfo {author} {\bibfnamefont {J.}~\bibnamefont {Powell}},
  \bibinfo {author} {\bibfnamefont {M.}~\bibnamefont {Cavagli{\`a}}}, \bibinfo
  {author} {\bibfnamefont {K.}~\bibnamefont {Ackley}}, \bibinfo {author}
  {\bibfnamefont {M.}~\bibnamefont {Bejger}}, \bibinfo {author} {\bibfnamefont
  {C.}~\bibnamefont {Chatterjee}}, \bibinfo {author} {\bibfnamefont
  {M.}~\bibnamefont {Coughlin}}, \bibinfo {author} {\bibfnamefont
  {S.}~\bibnamefont {Coughlin}}, \bibinfo {author} {\bibfnamefont
  {P.}~\bibnamefont {Easter}}, \bibinfo {author} {\bibfnamefont
  {R.}~\bibnamefont {Essick}}, \emph {et~al.},\ }\bibfield  {title} {\bibinfo
  {title} {Enhancing gravitational-wave science with machine learning},\
  }\href@noop {} {\bibfield  {journal} {\bibinfo  {journal} {Machine Learning:
  Science and Technology}\ }\textbf {\bibinfo {volume} {2}},\ \bibinfo {pages}
  {011002} (\bibinfo {year} {2020})}\BibitemShut {NoStop}%
\bibitem [{\citenamefont {Chatterji}\ \emph {et~al.}(2004)\citenamefont
  {Chatterji}, \citenamefont {Blackburn}, \citenamefont {Martin},\ and\
  \citenamefont {Katsavounidis}}]{Chatterji:2004qg}%
  \BibitemOpen
  \bibfield  {author} {\bibinfo {author} {\bibfnamefont {S.}~\bibnamefont
  {Chatterji}}, \bibinfo {author} {\bibfnamefont {L.}~\bibnamefont
  {Blackburn}}, \bibinfo {author} {\bibfnamefont {G.}~\bibnamefont {Martin}},\
  and\ \bibinfo {author} {\bibfnamefont {E.}~\bibnamefont {Katsavounidis}},\
  }\bibfield  {title} {\bibinfo {title} {{Multiresolution techniques for the
  detection of gravitational-wave bursts}},\ }\href
  {https://doi.org/10.1088/0264-9381/21/20/024} {\bibfield  {journal} {\bibinfo
   {journal} {Class. Quantum Grav.}\ }\textbf {\bibinfo {volume} {21}},\
  \bibinfo {pages} {S1809} (\bibinfo {year} {2004})},\ \Eprint
  {https://arxiv.org/abs/0412119} {arXiv:0412119 [gr-qc]} \BibitemShut
  {NoStop}%
\bibitem [{\citenamefont {Abbott}\ \emph
  {et~al.}(2017{\natexlab{b}})\citenamefont {Abbott} \emph
  {et~al.}}]{TheLIGOScientific:2017qsa}%
  \BibitemOpen
  \bibfield  {author} {\bibinfo {author} {\bibfnamefont {B.}~\bibnamefont
  {Abbott}} \emph {et~al.} (\bibinfo {collaboration} {LIGO Scientific
  Collaboration, Virgo Collaboration}),\ }\bibfield  {title} {\bibinfo {title}
  {{GW170817: Observation of Gravitational Waves from a Binary Neutron Star
  Inspiral}},\ }\href {https://doi.org/10.1103/PhysRevLett.119.161101}
  {\bibfield  {journal} {\bibinfo  {journal} {Phys. Rev. Lett.}\ }\textbf
  {\bibinfo {volume} {119}},\ \bibinfo {pages} {161101} (\bibinfo {year}
  {2017}{\natexlab{b}})},\ \Eprint {https://arxiv.org/abs/1710.05832}
  {arXiv:1710.05832 [gr-qc]} \BibitemShut {NoStop}%
\bibitem [{\citenamefont {Abbott}\ \emph
  {et~al.}(2021{\natexlab{b}})\citenamefont {Abbott} \emph
  {et~al.}}]{LIGOScientific:2021qlt}%
  \BibitemOpen
  \bibfield  {author} {\bibinfo {author} {\bibfnamefont {R.}~\bibnamefont
  {Abbott}} \emph {et~al.} (\bibinfo {collaboration} {LIGO Scientific, KAGRA,
  VIRGO}),\ }\bibfield  {title} {\bibinfo {title} {{Observation of
  Gravitational Waves from Two Neutron Star\textendash{}Black Hole
  Coalescences}},\ }\href {https://doi.org/10.3847/2041-8213/ac082e} {\bibfield
   {journal} {\bibinfo  {journal} {Astrophys. J. Lett.}\ }\textbf {\bibinfo
  {volume} {915}},\ \bibinfo {pages} {L5} (\bibinfo {year}
  {2021}{\natexlab{b}})},\ \Eprint {https://arxiv.org/abs/2106.15163}
  {arXiv:2106.15163 [astro-ph.HE]} \BibitemShut {NoStop}%
\bibitem [{\citenamefont {{Zackay}}\ \emph {et~al.}(2019)\citenamefont
  {{Zackay}}, \citenamefont {{Venumadhav}}, \citenamefont {{Roulet}},
  \citenamefont {{Dai}},\ and\ \citenamefont
  {{Zaldarriaga}}}]{2019arXiv190805644Z}%
  \BibitemOpen
  \bibfield  {author} {\bibinfo {author} {\bibfnamefont {B.}~\bibnamefont
  {{Zackay}}}, \bibinfo {author} {\bibfnamefont {T.}~\bibnamefont
  {{Venumadhav}}}, \bibinfo {author} {\bibfnamefont {J.}~\bibnamefont
  {{Roulet}}}, \bibinfo {author} {\bibfnamefont {L.}~\bibnamefont {{Dai}}},\
  and\ \bibinfo {author} {\bibfnamefont {M.}~\bibnamefont {{Zaldarriaga}}},\
  }\bibfield  {title} {\bibinfo {title} {{Detecting Gravitational Waves in Data
  with Non-Gaussian Noise}},\ }\href@noop {} {\bibfield  {journal} {\bibinfo
  {journal} {arXiv e-prints}\ ,\ \bibinfo {eid} {arXiv:1908.05644}} (\bibinfo
  {year} {2019})},\ \Eprint {https://arxiv.org/abs/1908.05644}
  {arXiv:1908.05644 [astro-ph.IM]} \BibitemShut {NoStop}%
\bibitem [{\citenamefont {Singer}\ and\ \citenamefont
  {Price}(2016)}]{Singer_2016}%
  \BibitemOpen
  \bibfield  {author} {\bibinfo {author} {\bibfnamefont {L.~P.}\ \bibnamefont
  {Singer}}\ and\ \bibinfo {author} {\bibfnamefont {L.~R.}\ \bibnamefont
  {Price}},\ }\bibfield  {title} {\bibinfo {title} {Rapid bayesian position
  reconstruction for gravitational-wave transients},\ }\bibfield  {journal}
  {\bibinfo  {journal} {Physical Review D}\ }\textbf {\bibinfo {volume} {93}},\
  \href {https://doi.org/10.1103/physrevd.93.024013}
  {10.1103/physrevd.93.024013} (\bibinfo {year} {2016})\BibitemShut {NoStop}%
\bibitem [{\citenamefont {Usman}\ \emph {et~al.}(2016)\citenamefont {Usman}
  \emph {et~al.}}]{Usman:2015kfa}%
  \BibitemOpen
  \bibfield  {author} {\bibinfo {author} {\bibfnamefont {S.~A.}\ \bibnamefont
  {Usman}} \emph {et~al.},\ }\bibfield  {title} {\bibinfo {title} {{The PyCBC
  search for gravitational waves from compact binary coalescence}},\ }\href
  {https://doi.org/10.1088/0264-9381/33/21/215004} {\bibfield  {journal}
  {\bibinfo  {journal} {Class. Quant. Grav.}\ }\textbf {\bibinfo {volume}
  {33}},\ \bibinfo {pages} {215004} (\bibinfo {year} {2016})},\ \Eprint
  {https://arxiv.org/abs/1508.02357} {arXiv:1508.02357 [gr-qc]} \BibitemShut
  {NoStop}%
\bibitem [{\citenamefont {Nitz}\ \emph {et~al.}(2018)\citenamefont {Nitz},
  \citenamefont {Dal~Canton}, \citenamefont {Davis},\ and\ \citenamefont
  {Reyes}}]{nitz2018rapid}%
  \BibitemOpen
  \bibfield  {author} {\bibinfo {author} {\bibfnamefont {A.~H.}\ \bibnamefont
  {Nitz}}, \bibinfo {author} {\bibfnamefont {T.}~\bibnamefont {Dal~Canton}},
  \bibinfo {author} {\bibfnamefont {D.}~\bibnamefont {Davis}},\ and\ \bibinfo
  {author} {\bibfnamefont {S.}~\bibnamefont {Reyes}},\ }\bibfield  {title}
  {\bibinfo {title} {Rapid detection of gravitational waves from compact binary
  mergers with pycbc live},\ }\href@noop {} {\bibfield  {journal} {\bibinfo
  {journal} {Physical Review D}\ }\textbf {\bibinfo {volume} {98}},\ \bibinfo
  {pages} {024050} (\bibinfo {year} {2018})}\BibitemShut {NoStop}%
\bibitem [{\citenamefont {Macleod}(2019)}]{gwpy}%
  \BibitemOpen
  \bibfield  {author} {\bibinfo {author} {\bibfnamefont {D.~e.~a.}\
  \bibnamefont {Macleod}},\ }\href@noop {} {\bibinfo {title} {Gwpy}},\ \bibinfo
  {howpublished} {\url{https://github.com/gwpy/gwpy}} (\bibinfo {year}
  {2019})\BibitemShut {NoStop}%
\bibitem [{\citenamefont {Allen}\ \emph {et~al.}(2012)\citenamefont {Allen},
  \citenamefont {Anderson}, \citenamefont {Brady}, \citenamefont {Brown},\ and\
  \citenamefont {Creighton}}]{Allen:2005fk}%
  \BibitemOpen
  \bibfield  {author} {\bibinfo {author} {\bibfnamefont {B.}~\bibnamefont
  {Allen}}, \bibinfo {author} {\bibfnamefont {W.~G.}\ \bibnamefont {Anderson}},
  \bibinfo {author} {\bibfnamefont {P.~R.}\ \bibnamefont {Brady}}, \bibinfo
  {author} {\bibfnamefont {D.~A.}\ \bibnamefont {Brown}},\ and\ \bibinfo
  {author} {\bibfnamefont {J.~D.~E.}\ \bibnamefont {Creighton}},\ }\bibfield
  {title} {\bibinfo {title} {{FINDCHIRP: An Algorithm for detection of
  gravitational waves from inspiraling compact binaries}},\ }\href
  {https://doi.org/10.1103/PhysRevD.85.122006} {\bibfield  {journal} {\bibinfo
  {journal} {Phys. Rev. D}\ }\textbf {\bibinfo {volume} {85}},\ \bibinfo
  {pages} {122006} (\bibinfo {year} {2012})},\ \Eprint
  {https://arxiv.org/abs/gr-qc/0509116} {arXiv:gr-qc/0509116} \BibitemShut
  {NoStop}%
\bibitem [{\citenamefont {Abbott}\ \emph {et~al.}(2016)\citenamefont {Abbott}
  \emph {et~al.}}]{LIGOScientific:2016aoc}%
  \BibitemOpen
  \bibfield  {author} {\bibinfo {author} {\bibfnamefont {B.~P.}\ \bibnamefont
  {Abbott}} \emph {et~al.} (\bibinfo {collaboration} {LIGO Scientific,
  Virgo}),\ }\bibfield  {title} {\bibinfo {title} {{Observation of
  Gravitational Waves from a Binary Black Hole Merger}},\ }\href
  {https://doi.org/10.1103/PhysRevLett.116.061102} {\bibfield  {journal}
  {\bibinfo  {journal} {Phys. Rev. Lett.}\ }\textbf {\bibinfo {volume} {116}},\
  \bibinfo {pages} {061102} (\bibinfo {year} {2016})},\ \Eprint
  {https://arxiv.org/abs/1602.03837} {arXiv:1602.03837 [gr-qc]} \BibitemShut
  {NoStop}%
\bibitem [{\citenamefont {Borbeau}(2018)}]{pycondor}%
  \BibitemOpen
  \bibfield  {author} {\bibinfo {author} {\bibfnamefont {J.}~\bibnamefont
  {Borbeau}},\ }\href@noop {} {\bibinfo {title} {Pycondor}},\ \bibinfo
  {howpublished} {\url{https://github.com/jrbourbeau/pycondor}} (\bibinfo
  {year} {2018})\BibitemShut {NoStop}%
\bibitem [{\citenamefont {Lo}(2021)}]{skymap-overlap}%
  \BibitemOpen
  \bibfield  {author} {\bibinfo {author} {\bibfnamefont {R.~K.~L.}\
  \bibnamefont {Lo}},\ }\href@noop {} {\bibinfo {title} {Skymap overlap}},\
  \bibinfo {howpublished}
  {\url{https://git.ligo.org/ka-lok.lo/skymap-overlap/}} (\bibinfo {year}
  {2021})\BibitemShut {NoStop}%
\bibitem [{\citenamefont {Boh\'e}\ \emph {et~al.}(2017)\citenamefont {Boh\'e}
  \emph {et~al.}}]{Bohe:2016gbl}%
  \BibitemOpen
  \bibfield  {author} {\bibinfo {author} {\bibfnamefont {A.}~\bibnamefont
  {Boh\'e}} \emph {et~al.},\ }\bibfield  {title} {\bibinfo {title} {{Improved
  effective-one-body model of spinning, nonprecessing binary black holes for
  the era of gravitational-wave astrophysics with advanced detectors}},\ }\href
  {https://doi.org/10.1103/PhysRevD.95.044028} {\bibfield  {journal} {\bibinfo
  {journal} {Phys. Rev. D}\ }\textbf {\bibinfo {volume} {95}},\ \bibinfo
  {pages} {044028} (\bibinfo {year} {2017})},\ \Eprint
  {https://arxiv.org/abs/1611.03703} {arXiv:1611.03703 [gr-qc]} \BibitemShut
  {NoStop}%
\bibitem [{\citenamefont {Coughlin}\ \emph {et~al.}(2020)\citenamefont
  {Coughlin} \emph {et~al.}}]{Coughlin:2020fwx}%
  \BibitemOpen
  \bibfield  {author} {\bibinfo {author} {\bibfnamefont {M.~W.}\ \bibnamefont
  {Coughlin}} \emph {et~al.},\ }\bibfield  {title} {\bibinfo {title}
  {{Implications of the search for optical counterparts during the second part
  of the Advanced LIGO\textquoteright{}s and Advanced Virgo\textquoteright{}s
  third observing run: lessons learned for future follow-up observations}},\
  }\href {https://doi.org/10.1093/mnras/staa1925} {\bibfield  {journal}
  {\bibinfo  {journal} {Mon. Not. Roy. Astron. Soc.}\ }\textbf {\bibinfo
  {volume} {497}},\ \bibinfo {pages} {1181} (\bibinfo {year} {2020})},\ \Eprint
  {https://arxiv.org/abs/2006.14756} {arXiv:2006.14756 [astro-ph.HE]}
  \BibitemShut {NoStop}%
\bibitem [{\citenamefont {Soni}\ \emph {et~al.}(2020)\citenamefont {Soni} \emph
  {et~al.}}]{LIGO:2020zwl}%
  \BibitemOpen
  \bibfield  {author} {\bibinfo {author} {\bibfnamefont {S.}~\bibnamefont
  {Soni}} \emph {et~al.} (\bibinfo {collaboration} {LIGO}),\ }\bibfield
  {title} {\bibinfo {title} {{Reducing scattered light in LIGO's third
  observing run}},\ }\href {https://doi.org/10.1088/1361-6382/abc906}
  {\bibfield  {journal} {\bibinfo  {journal} {Class. Quant. Grav.}\ }\textbf
  {\bibinfo {volume} {38}},\ \bibinfo {pages} {025016} (\bibinfo {year}
  {2020})},\ \Eprint {https://arxiv.org/abs/2007.14876} {arXiv:2007.14876
  [astro-ph.IM]} \BibitemShut {NoStop}%
\end{thebibliography}%
\end{document}